\documentclass[Original research]{elsarticle}

\usepackage{hyperref}
\usepackage{graphicx}
\usepackage{amsmath}
\usepackage{multirow}
\usepackage{tabularx}
\usepackage{threeparttable}   
\usepackage[justification=centering]{caption}
\usepackage{color}
\graphicspath{{figure/}}
\usepackage{algorithm}
\usepackage{algorithmic}

\journal{Journal of \LaTeX\ Templates}

\begin{document}

\begin{frontmatter}

\title{Super-resolving Compressed Images via Parallel and Series Integration of Artefacts Removal and Resolution Enhancement}

\author[aff1,aff2,aff3,aff31,aff4]{Hongming Luo}

\author[aff1,aff2,aff3,aff31,aff4]{Fei Zhou\corref{correspondingauthor}}
\cortext[correspondingauthor]{Corresponding author}
\ead{flying.zhou@163.com}

\author[aff1,aff2,aff3,aff31,aff4]{Guangsen Liao}
\author[aff1,aff3,aff4,aff6,aff7]{Guoping Qiu}

\address[aff1]{College of Electronics and Information Engineering, Shenzhen University, China}
\address[aff2]{Peng Cheng Laboratory, Shenzhen, China}
\address[aff3]{Guangdong Key Laboratory of Intelligent Information Processing, Shenzhen, China}
\address[aff31]{Shenzhen Key Laboratory of Digital Creative Technology, China}
\address[aff4]{Shenzhen Institute for Artificial Intelligence and Robotics for Society, Shenzhen, China}
\address[aff6]{School of Computer Science, University of Nottingham, Nottingham NG8 1BB, U.K.}
\address[aff7]{Guangdong-Hong Kong Joint Laboratory for Big Data Imaging and Communication, Shenzhen, Guangdong, China}

\begin{abstract}
In real-world applications, such as sharing photos on social media platforms, images are always not only sub-sampled but also heavily compressed thus often containing various artefacts. Simple methods for enhancing the resolution of such images will exacerbate the artefacts, rendering them visually objectionable. In spite of its high practical values, super-resolving compressed images is not well studied in the literature. In this paper, we propose a novel compressed image super resolution (CISR) framework based on parallel and series integration of artefacts removal and resolution enhancement. Based on a mathematical inference model for estimating a clean low-resolution (LR) image and a clean high-resolution (HR) image from a down-sampled and compressed observation, we have designed a CISR architecture consisting of two deep neural network modules: the artefacts removal module (ARM) and the resolution enhancement module (REM). The ARM and the REM work in parallel with both taking the compressed LR image as their inputs, at the same time they also work in series with the REM taking the output of the ARM as one of its inputs and the ARM taking the output of the REM as its other input. A technique called unfolding is introduced to recursively suppress the compression artefacts and restore the image resolution. A unique feature of our CISR system is that it exploits the parallel and series connections between the ARM and the REM, and recursive optimization to reduce the model’s dependency on specific types of degradation thus making it possible to train a single model to super-resolve images compressed by different methods to different qualities. Experiments are conducted on a mixture of JPEG and WebP compressed images without assuming apriori compression type and compression quality factor. To demonstrate our technique’s real-world application value, we have also applied the trained models directly to restore social media images which have undergone scaling and compression by unknown algorithms. Visual and quantitative comparisons demonstrate the superiority of our method over state-of-the-art super resolution methods, especially for heavily compressed images. Codes and datasets are available at https://github.com/luohongming/CISR\_PSI.git.
\end{abstract}


\end{frontmatter}


\section{Introduction}
\label{Intro}

Single image super resolution (SR) aims to reconstruct a high-resolution (HR) image from its low-resolution (LR) counterpart \cite{ref1}. It has received much attention due to its values in many applications, such as surveillance imaging \cite{ref3} and thumbnail image enlargement \cite{ref4}. In almost all practical applications, limited by storage capacity and transmission bandwidth, images are not only down-sampled but also compressed, especially in social media. If the compression is lossy, images are inevitably contaminated by annoying artefacts, such as blocking, ringing, fake edges, etc. Enhancing the resolution of such images will exacerbate the artefacts. Thus, in comparison to clean images, it is more challenging to super-resolve compressed images. Despite its practical application values, compressed image super resolution (CISR) is not well studied in the literature, and there are many problems remain unsolved. This work focuses on the single image SR (SISR) problem for compressed images.

\begin{figure}[!htb]
\begin{minipage}{0.4\linewidth}
\centerline{\includegraphics[scale=0.45]{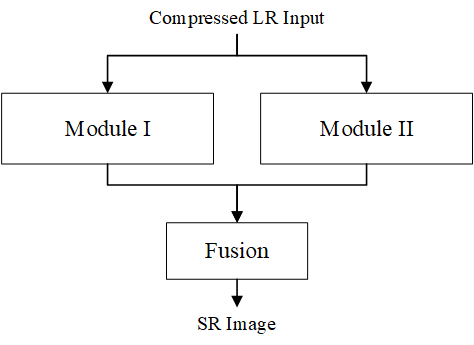}}
\centerline{(a)}
\end{minipage}
\hfill
\begin{minipage}{.6\linewidth}
  \centerline{\includegraphics[scale=0.45]{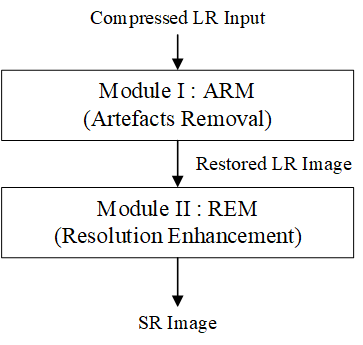}}
  \centerline{(b)}
\end{minipage}
\vfill
\begin{minipage}{1\linewidth}
  \centerline{\includegraphics[scale=0.45]{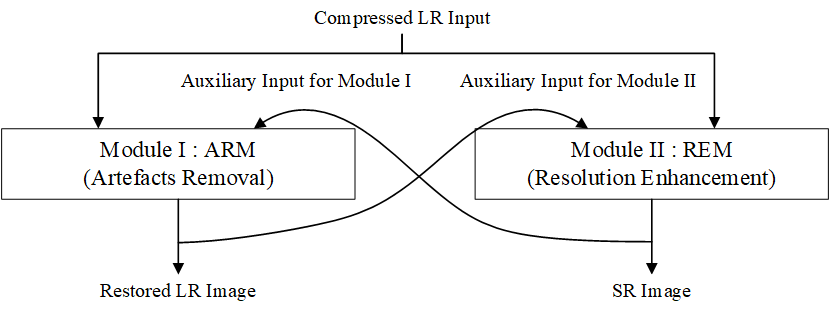}}
  \centerline{(c)}
\end{minipage}
\caption{Illustrations on different frameworks for CISR. (a) Parallel framework. (b) Series or cascaded framework. (c) Our framework.}
\label{fig:res1}
\end{figure}

To super-resolve compressed images, training a traditional SISR model with compressed data will have difficulty in producing high-quality super-resolved images (SRIs). The reason is that the resolution enhancement process will inevitably amplify high-frequency artefacts \cite{ref31}. In previous works, several specific models have been designed for CISR. They fall into two categories: joint model \cite{ref32, ref33} and series (or cascaded) model \cite{ref34, ref35}. Specifically, the joint model is a parallel architecture, where the input image or part of it streams to two independent modules simultaneously, and then the results from the two modules are fused to obtain the output, as shown in Fig. \ref{fig:res1}(a). However, two independent modules cannot benefit each other, limiting the performance of SISR for compressed images. In the cascaded model, the output of one module streams to the other module. Generally, two modules are involved: one is for compression artefacts removal and the other is for SR. If the SR module is applied first, artefacts would be amplified. It is much more difficult to suppress the amplified artefacts than the original signal. Therefore, in existing methods \cite{ref31, ref34, ref35}, the LR input image is first restored by reducing compression artefacts and then rescaled to a higher resolution, as shown in Fig. \ref{fig:res1}(b). However, some image details would be inevitably lost during the artefacts removal process, and those lost details will be difficult to be retrieved by subsequent modules.
Different from previous works, our framework is illustrated in Fig. \ref{fig:res1}(c). Our system is designed based on a mathematical inference for estimating a clean LR image and a clean HR image from a down-sampled and compressed observation. Formal derivation of our model design will be presented in Section \ref{overall}. A unique feature of our method is that it exploits the parallel and series connections between the ARM and the REM, and recursive optimization to reduce the model’s dependency on specific types of degradation thus making it possible to train a single model to super-resolve images compressed by different methods to different qualities. This is important, as in many real-world applications such as photo sharing on social media platforms, images would have always undergone scaling and compression by unknown algorithms. 

As shown in Fig. \ref{fig:res1}(c), our method consists of two modules, the ARM (Module I) and the REM (Module II). However, these two modules are not simply parallel or cascaded. On one hand, the compressed LR input streams to the two modules in a parallel way. On the other hand, the output of one module is fed back to the other module, resulting in two series flows. Here, we regard the output of one module as the auxiliary input of the other one. Essentially, both parallel and series flows are involved in our framework. Its advantages are threefold: First, the original information in the input is fully available to the both modules. Second, the output of Module I facilitates Module II by providing it with a relatively clean version of the LR image. Third, the output of Module II supplies Module I with an image with high-frequency details which are frequently lost during the artefacts removal process.

In this work, the both modules are implemented by deep neural networks, and both training and testing are achieved by a recursive process sometimes referred to as unfolding \cite{ref36}. In addition, at the input end of each module, we include a modified non-local operator to capture long-range dependencies in images. By using this modified non-local operator, a relatively clean but blurry image named non-locally filtered image is provided. In an SISR deep network, a long skip connection has been demonstrated to be highly effective in forcing the network to learn residuals, \textit{i.e.}, high-frequency image details \cite{ref15}. It not only allows low-frequency information to take a shortcut, but also alleviates the problem of vanishing or exploding gradients \cite{ref37}. Hence, a long skip connection is also adopted in each module of our framework. Since three images, \textit{i.e.}, the input, the auxiliary input, and the non-locally filtered image, are available, we propose to adaptively combine them for the skip connection by learning their respective weights during training. That is, all three images are connected to the output with learnable contributions.

To demonstrate the effectiveness of the proposed framework, we collect a photograph dataset and a \textit{WeChat} avatar image dataset. The photograph dataset is used to generate compressed images by different compression methods with various quality factors. The \textit{WeChat} avatar image dataset contains LR images that have undergone compression and scaling by \textit{WeChat}’s internal algorithms (unknown to users). More details about these two datasets are provided in Section \ref{dataset}. 
The main contributions of this work are as follows:

\begin{itemize}

\item We have developed a new framework for compressed image super resolution (CISR). Our method consists of an artefacts removal module (ARM) and a resolution enhancement module (REM) which are connected in parallel and in series. A unique feature of our CISR system is that it exploits the parallel and series connections between the ARM and the REM, and recursive optimization to reduce the model’s dependency on specific types of degradation thus making it possible to train a single model to super-resolve images compressed by different methods to different qualities.
\item We present two datasets which would benefit research in super-resolving compressed images. One dataset contains photography images compressed by two most widely-used compression methods JPEG and WebP, and the other dataset contains real-world images that has undergone compression and scaling by unknown algorithms from one of the world’s largest social media platform \textit{WeChat}.
\item We present extensive experimental results to demonstrate that our new method outperforms state-of-the-art based on quantitative measures and visual comparison.
\end{itemize}

\section{Related Work}
\label{relatedwork}

\subsection{Super Resolution}
\label{SISR}
Early models in SISR are example-based \cite{ref5}, where a search for the nearest neighbor is performed with compatibility constraints. Subsequent famous methods include the models based on locally linear embedding \cite{ref7}, sparse coding \cite{ref8}, and neighborhood regression \cite{ref9}, etc. All of these various techniques, called traditional methods, are limited by their shallow architectures. One can refer to \cite{ref1} for the detailed review of traditional SISR methods.

After the first successful attempts to adopt deep networks in SISR tasks \cite{ref13}, many powerful techniques emerge to make deep networks more effective in the SISR task. For example, residual learning is introduced to achieve a very deep network. Successful applications of residual learning in SR include: global residual learning in \cite{ref15}, local residual learning in enhanced deep super-resolution network (EDSR) \cite{ref16}. Attention mechanism is also considered in the residual channel attention network (RCAN) \cite{ref19} and second-order attention network (SAN) \cite{ref20}. In feedback network \cite{ref24}, deep networks are unfolded to make their training feasible. The success of unfolding technique in deep SISR models motivates us to employ it to train our model in Fig. \ref{fig:res1}(c), where the auxiliary inputs can be treated as feedback. Moreover, non-local means has be successfully adopted in both traditional and deep SISR methods. A non-local total variation prior \cite{ref25} are applied in the traditional SISR framework. In deep SISR models \cite{ref26}, non-local means is known as the spatial attention that is essentially a non-local convolution process. One can refer to \cite{ref11} for comprehensive surveys of deep SISR models.

Albeit great success of the above methods for clean images, they fail to super-resolve the images with multiple degradations. To reduce the simulated-to-real gap \cite{ref28}, some SISR methods are developed for the images with various degradations.

To super-resolve noisy LR images, the method in \cite{ref29} combines a noisy SR image and a de-noised SR image. In \cite{ref39}, the noise level of LR images is estimated to determine the value of regularization parameter. Recently, the noise-robust iterative back-projection (NRIBP) is presented in \cite{ref40} for noisy image SR. In addition to noises, some models further consider the impact of various blur kernels in SISR. In \cite{ref41}, blur kernel and noise level are regarded as one of the inputs, making deep networks possible to handle blurring and noises. In \cite{ref42}, an auxiliary variable is introduced to separate the problem of blurred image SR to two iterative sub-problems, \textit{i.e.}, image restoration problem and SR problem. And its extension named unfolding super-resolution network (USRNet) \cite{ref43} introduces a trainable prior module. With regard to the unknown blur kernel, kernel prediction \cite{ref30} method and generative adversarial networks \cite{ref45} are used in SR framework. However, it is required that degradation parameters are available in most methods mentioned above, \textit{e.g.}, \cite{ref41, ref42, ref43, ref44}. 
In these above literatures, Gaussian noises and blurring are most considered. However, compression-induced artefacts are totally different from Gaussian noises and blurring. First, Gaussian noises are independent from image contents, and the noise distribution generally remains the same over the whole image. In contrast to Gaussian noises, compression artefacts would be highly related to image contents and spatially variant. Second, blurring kernels can be spatially variant but produce no high-frequency artefacts. Different from blurring, compression could give rise to high-frequency artefacts, \textit{e.g.}, blocking. Therefore, the models designed for noisy and blurred images are not appropriate for compressed images, as we will demonstrate in Section \ref{compare}.

In comparison with noisy and blurred image SR, studies on the SISR problem for compressed images are relatively seldom. The method in \cite{ref31} performs an iterative regularization and SR procedure. In \cite{ref32}, image patches are classified into two sets of blocking and non-blocking to super-resolve them separately. Regarding compression artefacts as noises, the method in \cite{ref33} adopts a similar strategy in \cite{ref29}. However, the methods in \cite{ref31}–\cite{ref33} ignore the information exchange between different modules. The model of iterative cascaded SR and de-blocking (ICSD) \cite{ref34} is the first attempt to use the information exchange between de-blocking and SR. In \cite{ref35}, CISR is implemented by deep convolution neural networks (CISRDCNN), which consists of three cascaded modules to obtain SR images. However, the losses of details in previous modules can hardly be retrieved by the subsequent ones. Moreover, the information exchange among different modules is not fully exploited. 

\subsection{Compression Artefacts Removal}
\label{car}
Traditional methods of compression artefacts removal can be performed in either spatial domain \cite{ref46, ref47} or transform domain \cite{ref48, ref49}. The method of shape-adaptive DCT (SA-DCT) \cite{ref49} defines the shape of the transform supporting in a point-wise adaptive way to produce clean edges. Similar to the development of SISR, deep learning has also achieved success in the field of compression artefacts removal. The pioneer work in \cite{ref50} first introduces convolutional network to achieve compression artefacts removal. Subsequently, more deep models are presented, such as residual learning in de-noising convolutional neural network (DnCNN) \cite{ref53} and deep convolutional sparse coding (DCSC) model \cite{ref55}, etc. One can refer to \cite{ref56} for the detailed review of studies on compression artefacts removal. 

It would be interesting to see that many techniques or models have been successfully applied to both the problems of SISR and compression artefacts removal. For example, the trainable nonlinear reaction diffusion (TNRD) model can be trained with different reaction terms to solve the two problems \cite{ref57}. As well-known techniques in SISR, sparse coding \cite{ref55, ref58}, non-local means \cite{ref59} and attention networks \cite{ref60} are also employed to reduce compression artefacts. This phenomenon can be attributed to the common properties of reproduced images and learning models. 

\section{Proposed SISR Model for Compressed Images}

\subsection{Overall framework}
\label{overall}
Before discussing the problem of super-resolving compressed images, we review the SR of clean images. Let us consider a clean LR image \textbf{y} and its corresponding HR counterpart \textbf{x}. Their relation can be formulated as
\begin{equation}
\mathbf{y} = \mathbf{Dx},
\end{equation}
where \textbf{D} is a sub-sampling operator. SR seeks to reverse the sub-sampling procedure in Eq. (1) and find a mapping $\mathcal{F}: \mathbf{y} \rightarrow \mathbf{x}$. Recently, many researchers model this mapping as a deep convolutional neural network (DCNN), such as \cite{ref13}, \cite{ref16}, \cite{ref18}, \cite{ref80} etc. Therefore, these models can be represented as 
\begin{equation}
 \hat{\mathbf{x}} = \mathcal{F}(\mathbf{y};\Theta_{\mathcal{F}}),
\end{equation}
where $\Theta_{\mathcal{F}}$ is the parameter set of DCNN. They directly learn from a training set of degraded and ground-truth image pairs by an end-to-end training.

Back to the problem of super-resolving compressed images, the biggest difference from the traditional SR problem is that compression is involved. The compression procedure can be formulated by
\begin{equation}
	\mathbf{z} = \mathbf{T}^{-1}\mathbf{QTy},
\end{equation}
where \textbf{z} denotes a compressed LR image, \textbf{T} is a linear transform used in compression, $ \mathbf{T}^{-1} $ is the corresponding inverse transform, and \textbf{Q} is a quantization operator. In this work, we do not assume any specific transform, although the discrete cosine transform (DCT) is widely used, \textit{e.g.}, in JPEG standard \cite{ref61}. The relation between \textbf{z} and \textbf{x} can be given by  
\begin{equation}
	\mathbf{z} = \mathbf{Cx},
\end{equation}
where $\mathbf{C} = \mathbf{T}^{-1}\mathbf{QTD}$. From Eqs. (1) and (4), the clean HR image \textbf{x} can be reconstructed from either its clean LR version \textbf{y} or its compressed LR version \textbf{z}. Therefore, a mapping function that uses both \textbf{y} and \textbf{z} to restore \textbf{x} is desirable, \textit{i.e.}, we would like a mapping $\mathcal{R}: (\mathbf{y}, \mathbf{z}) \rightarrow \mathbf{x}$. Assuming both \textbf{y} and \textbf{z} are available, similar to the SR of clean images in Eq. (2), we can model the mapping function $\mathcal{R}$ as
\begin{equation}
\hat{\mathbf{x}} = \mathcal{R}(\mathbf{y}, \mathbf{z};\Theta_{\mathcal{R}}),
\end{equation}
where $\Theta_{\mathcal{R}}$ represents the parameter set of $\mathcal{R}$. Notice that Eq. (5) has two inputs, \textit{i.e.}, \textbf{y} and \textbf{z}, but only \textbf{z} is available in our problem. Therefore, using Eq. (5) to estimate \textbf{x}, it is necessary to recover \textbf{y}. 

In Eq. (1), the clean LR image \textbf{y} can be obtained by the sub-sampling procedure from the HR image \textbf{x} while in Eq. (3), \textbf{y} can be generated by the compressed image restoration procedure from the compressed LR image \textbf{z}. Hence, a mapping that takes both \textbf{x} and \textbf{z} as input to estimate \textbf{y} is desirable, \textit{i.e.}, we require $\mathcal{P}: (\mathbf{x}, \mathbf{z}) \rightarrow \mathbf{y}$. Similarly, assuming both \textbf{x} and \textbf{z} are available, we can model the mapping function $\mathcal{P}$ as 
\begin{equation}
	\hat{\mathbf{y}} = \mathcal{P}(\mathbf{x}, \mathbf{z};\Theta_{\mathcal{P}}),
\end{equation}
where $\Theta_{\mathcal{P}}$ represents the parameter set of $\mathcal{P}$.

Based on Eqs. (5) and (6), we propose a novel framework that integrates both parallel model and series model as shown in Fig. \ref{fig:res1}(c). The artefacts removal module (ARM), \textit{i.e.}, Module I is derived from Eq. (6) while the resolution enhancement module (REM), \textit{i.e.}, Module II is associated with Eq. (5). Specifically, for the two inputs in Eq. (6), one is the compressed LR input image \textbf{z}, while the other (which is called auxiliary input) is the clean HR image \textbf{x}. Note that in the training stage, \textbf{x} is available as the training target but not available in the testing stage, therefore we have to use the output of module II to replace \textbf{x} in Eq. (6). Similarly, for the two inputs in Eq. (5), one is the compressed LR input image \textbf{z}, while the other is the clean LR image \textbf{y}. Again in the training stage \textbf{y} is available but it is not available in the testing stage, it will therefore have to come from the output of module I instead, \textit{i.e.}, an estimated clean version of \textbf{z} which replace \textbf{y} in Eq. (5). To solve Eqs. (5) and (6), we utilize the strategy of recursive optimization. Since the estimation of \textbf{y} is easier than \textbf{x}, we perform the estimation of \textbf{y} first. That is, we generate the output of the ARM module using $\hat{\mathbf{x}}$, then obtain the output of the REM module using $\hat{\mathbf{y}}$ recursively. Hence, the whole recursive procedure can be rewritten to 
\begin{equation}
\begin{aligned}
&\hat{\mathbf{y}}_{j} = \mathcal{P}(\hat{\mathbf{x}}_{j-1}, \mathbf{z}; \Theta_{\mathcal{P}}), \\
&\hat{\mathbf{x}}_{j} = \mathcal{R}(\hat{\mathbf{y}}_{j}, \mathbf{z}; \Theta_{\mathcal{R}}), \\
\end{aligned}
\end{equation}
where $j \leq J $ indicates the index of iteration, $J$ is a preset maximum iteration number. $\hat{\mathbf{y}}_{j}$ and $\hat{\mathbf{x}}_{j}$ are the estimation of \textbf{y} and \textbf{x} at \textit{j-th} iteration respectively. This strategy is also known as deep unfolding or unrolling in the field of deep learning \cite{ref36}.
\begin{figure*}[!htb]
\centerline{\includegraphics[scale=0.3]{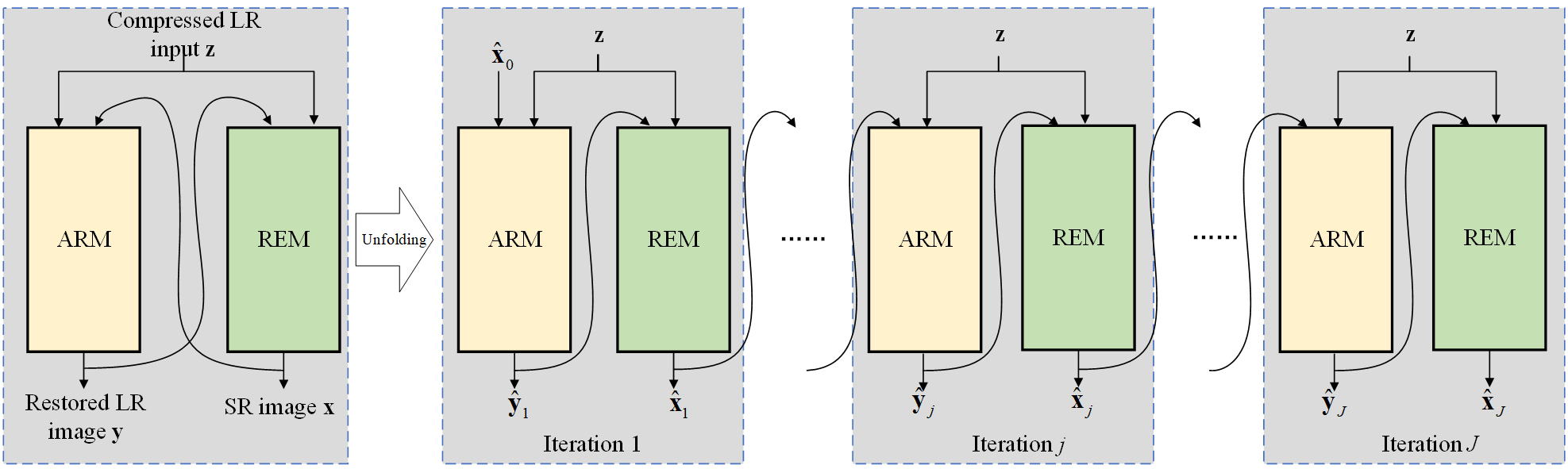}}
\caption{Overall framework of the proposed model.}
\label{fig:res2}
\end{figure*} 

The unfolding of our parallel and series integration framework is illustrated in Fig. \ref{fig:res2}. \textbf{From Eq. (7), one can see that, there are only two mapping functions in the recursive optimization. That is, the parameters $\Theta_{\mathcal{P}}$ in the ARM module are shared in each iteration, and the parameters $\Theta_{\mathcal{R}}$ in the REM module are also shared across different iteration indexes.} Each dash box in Fig. \ref{fig:res2} represents an iteration and there are two outputs in each iteration, \textit{i.e.}, $\hat{\mathbf{y}}$ and $\hat{\mathbf{x}}$. The subscript $j$ in $\hat{\mathbf{x}}_{j}$ is the same as the $j$ in Eq. (7). These two outputs in each iteration are required in the loss function for model training and we calculate the loss for every iteration. Given a $N$ training samples $\{ \mathbf{z}^{(i)}, \mathbf{y}^{(i)}, \mathbf{x}^{(i)}\}_{i=1}^{N}$, the loss function $\mathcal{L}$ can be written as 
\begin{equation}
\begin{aligned}
\mathcal{L}(\Theta_{\mathcal{P}}, \Theta_{\mathcal{R}})=& \frac{1}{NJ}\sum_{i=1}^{N}\sum_{j=1}^{J}\rho_{j}({||\mathcal{P}(\hat{\mathbf{x}}_{j-1}^{(i)},\mathbf{z}^{(i)};\Theta_{\mathcal{P}})-\mathbf{y}^{(i)}||}_{1} \\
&+ \gamma {||\mathcal{R}(\hat{\mathbf{y}}_{j}^{(i)},\mathbf{z}^{(i)};\Theta_{\mathcal{R}})-\mathbf{x}^{(i)}||}_{1}), \\
\end{aligned}
\end{equation}
where $\rho_{j}$ controls loss weight of each iteration, and $\gamma$ balances the impacts of the ARM and the REM. According to curriculum learning strategy \cite{ref68}, we regard the training in the first few iterations as easy tasks by setting smaller loss weight $\rho_{j}$ for smaller $j$. Using the unfolding technique and the loss defined in Eq. (8), an end-to-end training is performed to fix the parameters $\Theta_{\mathcal{P}}$ and $\Theta_{\mathcal{R}}$ in our model. After training the model, for a given LR compressed image, the proposed model can produce $J$ results for $\hat{\mathbf{x}}$. It is important to note that unlike conventional memory-less feedforward neural network architecture, our system is recursive in both the training and testing stages. In the training stage,  training samples have to go through $J$ iterations, each iteration produces two outputs which are compared with the ground truths for training. Similarly, in the testing stage, an input image also has to go through $J$ iterations, each produces a version of the super-resolved image with increasing accuracy, and the version from the final iteration which should contain the most details is normally used as the final output. In order to set up the whole recursive optimization procedure, the initial estimation of \textbf{x}, \textit{i.e.}, $\hat{\mathbf{x}}_{0}$, is obtained by bicubicly up-samling \textbf{z}. Details of the training and testing stages are shown in pseudo code in Algorithm 1 and Algorithm 2. 
\begin{algorithm}
	\caption{Training stage}
	\label{alg:1}
	\begin{algorithmic}[1]
		\REQUIRE Distribution of HR images $p(X)$. 
		\STATE Initialize the parameters of the ARM and the REM $\Theta_{\mathcal{P}}$, $\Theta_{\mathcal{R}}$ and set the maximum iteration number as $J$ and the sampling size of images as $N$.
		\REPEAT 
		\STATE Sample a batch of images $\{\mathbf{x}^{(i)} \}_{i=1}^{N} \sim p(X)$.
		\STATE Generate clean LR images $\{\mathbf{y}^{(i)}\}_{i=1}^{N}$ and compressed LR images $\{\mathbf{z}^{(i)}\}_{i=1}^{N}$ from $\{\mathbf{x}^{(i)}\}_{i=1}^{N}$.
		\STATE Set $\{\hat{\mathbf{x}}_{0}^{(i)}\}_{i=1}^{N}$ as bicubicly up-sampled $\{\mathbf{z}^{(i)}\}_{i=1}^{N}$.
		\FOR {$i=1$ to $n$}
		\FOR {$j=1$ to $J$}
		\STATE $\hat{\mathbf{y}}_{j}^{(i)} = \mathcal{P}(\hat{\mathbf{x}}_{j-1}^{(i)}, \mathbf{z}^{(i)}; \Theta_{\mathcal{P}})$,
		\STATE $\hat{\mathbf{x}}_{j}^{(i)} = \mathcal{R}(\hat{\mathbf{y}}_{j}^{(i)}, \mathbf{z}^{(i)}; \Theta_{\mathcal{R}})$,
		\ENDFOR 
		\ENDFOR
		\STATE Compute the gradients $\nabla_{\Theta_{\mathcal{P}}}\mathcal{L}$ and $\nabla_{\Theta_{\mathcal{R}}}\mathcal{L}$ using $\mathcal{L}$ in Eq. (8).
		\STATE Update the parameters $\Theta_{\mathcal{P}}$ and $\Theta_{\mathcal{P}}$ using $\nabla_{\Theta_{\mathcal{P}}}\mathcal{L}$ and $\nabla_{\Theta_{\mathcal{R}}}\mathcal{L}$ respectively.
		\UNTIL{convergence}
\end{algorithmic}
\end{algorithm} 
\begin{algorithm}
	\caption{Testing stage}
	\label{alg:1}
	\begin{algorithmic}[1]
		\REQUIRE Compressed LR image $\mathbf{z}$.
		\REQUIRE Parameters of trained model $\Theta_{\mathcal{P}}$, $\Theta_{\mathcal{R}}$ and the maximum iteration number $J$.
		\STATE Set $\hat{\mathbf{x}}_{0}$ as bicubicly up-sampled \textbf{z}.
		\FOR {$j=1$ to $J$}
		\STATE $\hat{\mathbf{y}}_{j} = \mathcal{P}(\hat{\mathbf{x}}_{j-1}, \mathbf{z}; \Theta_{\mathcal{P}})$
		\STATE $\hat{\mathbf{x}}_{j} = \mathcal{R}(\hat{\mathbf{y}}_{j}, \mathbf{z}; \Theta_{\mathcal{R}})$
		\ENDFOR 
		\STATE Take $\hat{\mathbf{x}}_{J}$ as the final estimation of HR image.
\end{algorithmic}
\end{algorithm} 

Obviously, the estimations of both \textbf{x} and \textbf{y} will not be perfect. False patterns are very likely to appear and will be further amplified by the sequential processing. These false patterns would cause the results to diverge from real image contents. From this perspective, it is essential to use the original compressed LR image \textbf{z} as an input in each modules during the unfolding process. The input \textbf{z} helps reducing the impact of accumulated errors on the two modules by continuously providing each module with the original signals. 

As we will show in the experiment (Section \ref{diff_compress}), a distinctive advantage of our model is that a single trained model can better handle input image compressed to different qualities than either the series or the parallel architecture. It is not difficult to understand this property from the design of the model. From Eq. (5), we can see that the super-resolved output has a clean input image \textbf{y}. Although \textbf{y} is not available in practice, we use a version of \textbf{y} restored from the compressed input \textbf{z}, \textit{i.e.}, $\hat{\mathbf{y}}$ in equation Eq. (6), as a substitute of the clean \textbf{y}. In this way, we have reduced the impact of compression on the final super-resolved results. It is also worth noting that Eq. (6) takes both \textbf{x} and \textbf{z} as input (although in practice an estimated version of \textbf{x}, \textit{i.e.}, $\hat{\mathbf{x}}$ in Eq. (5) is used), this input has provided the ARM with high-frequency details which will prevent $\hat{\mathbf{y}}$ from becoming excessively blur, thus improving the quality of $\hat{\mathbf{y}}$.

\subsection{Architectures of Modules I and II}
\label{arch_module}
\begin{figure*}[ht]
\begin{minipage}{1\linewidth}
\centerline{\includegraphics[scale=0.4]{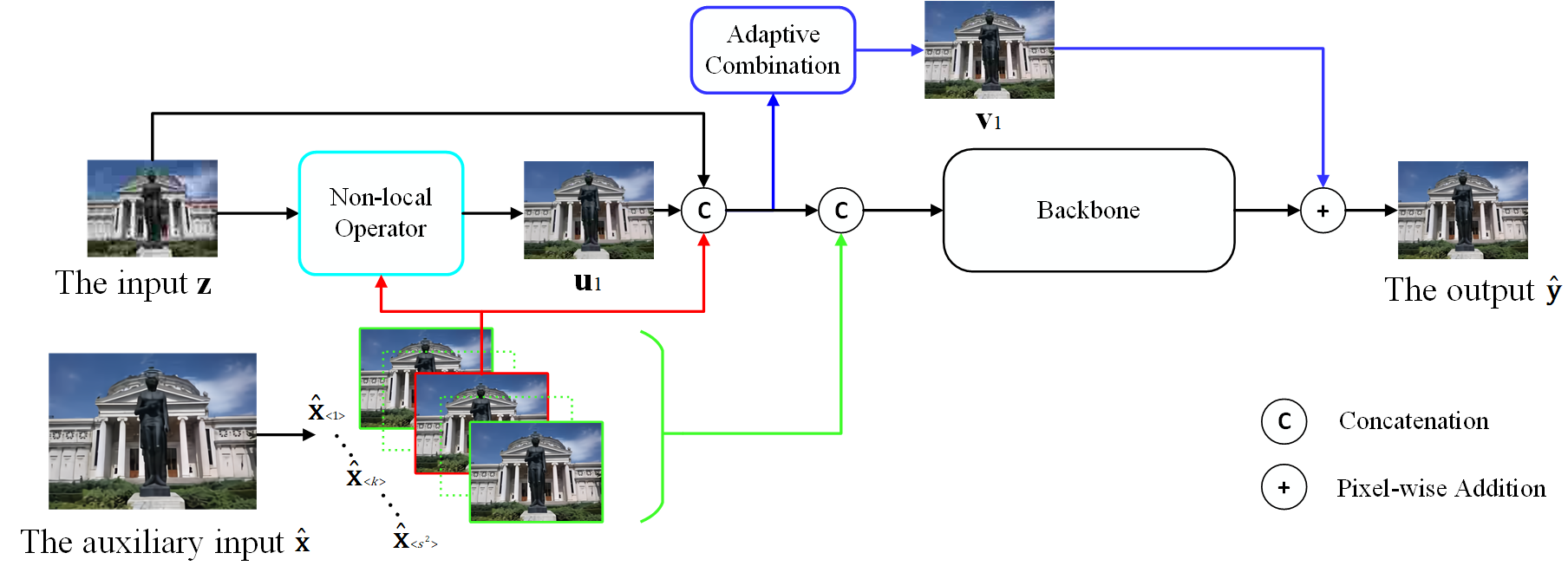}}
\centerline{(a)}
\end{minipage}
\begin{minipage}{1\linewidth}
  \centerline{\includegraphics[scale=0.4]{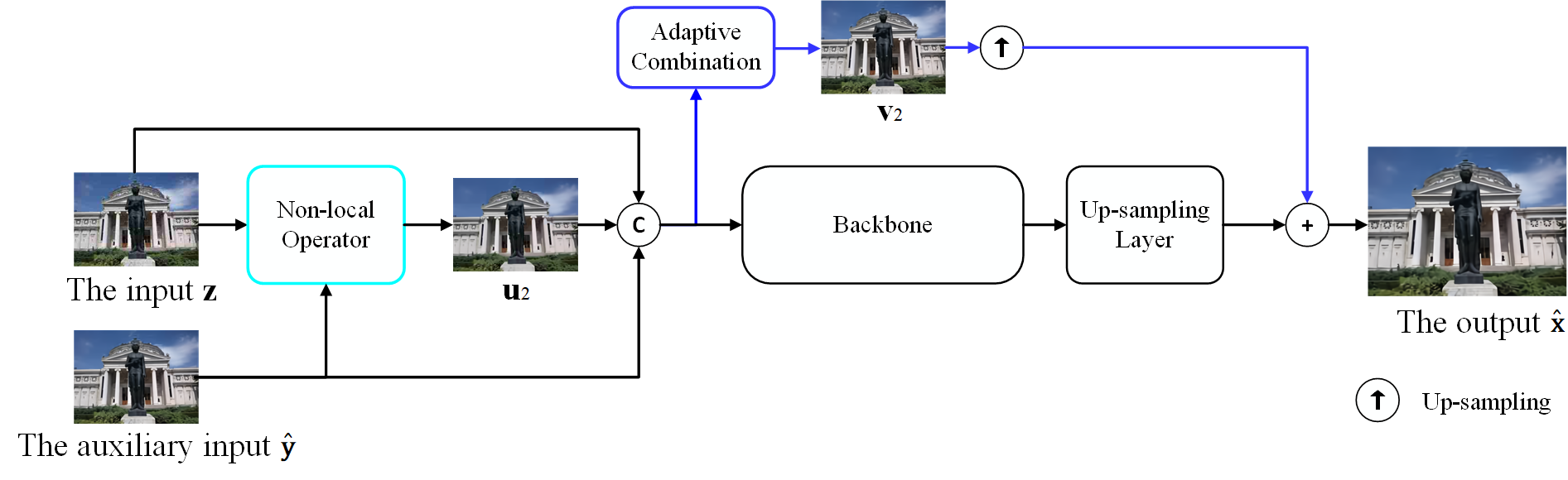}}
  \centerline{(b)}
\end{minipage}
\caption{Architectures of the two modules in our framework. (a) Module I: ARM. (b) Module II: REM.}
\label{fig:res3}
\end{figure*}

In this sub-section, we will detail the mappings $\mathcal{P}(\cdot)$ and $\mathcal{R}(\cdot)$, which are implemented via DCNN. The mapping $\mathcal{P}(\cdot)$ in Eq. (7) essentially achieves artefacts removal, and the mapping $\mathcal{R}(\cdot)$ in Eq. (7) is used for the resolution enhancement. As mentioned in Section \ref{car}, the solutions to these two problems share many common techniques and models. Therefore, we employ similar architectures for both mappings, as illustrated in Fig. \ref{fig:res3}. Most recently-developed architectures and related techniques can be used to embody the ARM and the REM. In this work, we adopt several residual groups of RCAN \cite{ref19} as the backbone of these two modules. Moreover, in each module, we include a non-local operator, highlighted in cyan in Fig. \ref{fig:res3}. The non-local operator can benefit the mappings in Eq. (7) by capturing long-range dependencies over the whole image. Different from the previous non-local operator that only operates on the input image, we make use of both the original input and the auxiliary input to weaken the influence of blocking artefacts. After the non-local operator, its output, the original input, and the auxiliary input are concatenated for subsequent processing. A long skip connection, marked in blue in Fig. \ref{fig:res3}, is used to pass the concatenated images to the output by a shortcut. Thereinto, an adaptive combination of concatenated images is exploited to fuse the pass-by information from different sources for the long skip connection. 
It is worth noting that the architectures of the ARM and the REM are not exactly the same. The reason is that the resolutions of the auxiliary inputs and the outputs are different for the two modules. In Module II, the resolution of the auxiliary input $\hat{\mathbf{y}}$ is the same as that of the input \textbf{z}. Whereas, in Module I, the width and height of the auxiliary input $\hat{\mathbf{x}}$ are \textit{s} times larger than those of \textbf{z}, where \textit{s} is the upscale factor. A straightforward idea is to down-scale $\hat{\mathbf{x}}$ to the same resolution as \textbf{z}, just like the strategy in \cite{ref34}. However, many important image details would be lost from $\hat{\mathbf{x}}$ after down-scaling. Hence, to keep the full information of $\hat{\mathbf{x}}$, we rearrange it into $\mathit{s}^{2}$ copies by using the space-to-depth transformation in \cite{ref63}, which can be regarded as the inverse process of pixel shuffle. All the copies, denoted as $\hat{\mathbf{x}}_{<1>}, \hat{\mathbf{x}}_{<2>}, …, \hat{\mathbf{x}}_{<\mathit{s}^{2}>},$ have similar image contents and the same resolution as the input \textbf{z}. Among those copies, there exist sub-pixel displacements whose values are the multiples of 1/\textit{s}. According to the space-to-depth operator, the copy $\hat{\mathbf{x}}_{<\mathit{k}>}$ is well registered with \textbf{z}, where $\mathit{k}$ equals to the number rounding off $(\mathit{s}^{2}+1)/2$. Thus, this copy, highlighted in red in Fig. \ref{fig:res3}(a), is fed to the non-local operator and the adaptive combination for long skip connection. The remaining copies, marked in green in Fig. \ref{fig:res3}(a), are also concatenated with the pass-by information and then inputted into the backbone. Another difference between the ARM and the REM lies in the output end. In the ARM, the output is of the same resolution as the input \textbf{z}, while the output resolution in the REM is $\mathit{s}^{2}$ times larger than the input resolution. Thus, after the backbone in the REM we deploy an up-sampling layer, which consists of $\mathit{s}^{2}$ convolutional filters followed by an operation of pixel shuffle. Correspondingly, a simple up-sampling operator, such as bicubic interpolation, is adopted in the long skip connection.

In the followings, we will provide the details in the modified non-local operator and the adaptive combination, respectively.

\subsubsection{Modified non-local operator}
\label{nonlocal}

The idea of non-local means is to utilize the self-similarity of images \cite{ref64, ref65}. Similar patches or features from long-range positions are selected as candidates to recover local signals which may be lost due to artefacts or down-sampling. The non-local operator performed on the input \textbf{z} can be defined as
\begin{equation}
\mathbf{u}_{\mathit{a}}=\sum_{n} \mathbf{w}(m,n)\cdot\mathbf{z}(n),
\end{equation}
where $\mathbf{u}_{\mathit{a}}$ is the output of non-local operator, \textit{a} is 1 for Module I and 2 for Module II, $m$ is the index of local patch to be recovered, $n$ is the index of candidate patches over the whole image, and \textbf{w} is the similarity matrix of \textbf{z}. In Eq. (9), all the candidates are directly from \textbf{z} to prevent any processing or manipulation on the original signal.

For a non-local operator, the measure of similarity is very important. When measuring the similarity in compressed images, we should be aware of the high similarity among the patches containing blocking artefacts. The horizontal or vertical signal patterns of blocking patches are highly similar to each other and may repeatedly appear in compressed images. Given a blocking patch, we wish to employ the patches that are similar in image content, instead of the blocking pattern. Therefore, we resort to the auxiliary input to measure the similarity matrix \textbf{w}. Specifically, the similarity between the \textit{m-th} patch and the \textit{n-th} one is calculated as
\begin{equation}
\mathbf{w}(m,n) =\frac{1}{S} \cdot exp(-\frac{{||\mathbf{g}(m)-\mathbf{g}(n)}||^{2}}{\mathbf{h}(m)^{2}}) \cdot d(\mathbf{z}(n)),
\end{equation}
where $S$ is the normalized parameter to make the summation of each row in \textbf{w} equals to 1, \textbf{h} is an adaptive parameter map which will be explained later, and \textbf{g} denotes an image from the auxiliary input. In Module I, \textbf{g} is selected as $\hat{\mathbf{x}}_{<k>}$, and it is the auxiliary input $\hat{\mathbf{y}}$ in Module II. Thus, \textbf{g} has exactly the same size as the input \textbf{z} in the both modules. In Eq. (10), we further include a binary function $d(\cdot)$ to detect the blocking edges in \textbf{z}. This function returns $0$ for blocking patches and $1$ otherwise. In this work, we adopt the method in \cite{ref66} to implement the detection function $d(\cdot)$. Both \textbf{g} and $d(\cdot)$ are essential to calculate the similarity matrix \textbf{w}. By using \textbf{g} instead of \textbf{z} to measure the patch similarity, we can pay attention to the candidate patches with similar contents rather than similar blocking patterns. By using the binary function $d(\cdot)$, we can discard the blocking patches of \textbf{z} in the calculation of Eq. (9). It means that some candidates with similar contents will be excluded if they are severely contaminated by blocking artefacts in \textbf{z}.

In the measure of the similarity matrix, the role of parameters \textbf{h} is to control the sparsity of the similarity matrix. Generally, a large value of the parameter would result in a smooth result, while a small one would produce artefacts and noises in $\mathbf{u}_{a}$. It has been demonstrated that the selection of \textbf{h} has great impact on the results of non-local operator \cite{ref59}. Moreover, its selection should depend on image content. For smooth regions, a large value of $\mathbf{h}$ is preferred. For textural regions, the reverse applies. Therefore, in this work, we employ a simple DCNN to adaptively estimate \textbf{h} from \textbf{g}. This network only consists of two convolutional layers with a layer of rectified linear unit (ReLU) between them. Since the parameter map \textbf{h} has the same size as \textbf{g}, we get a pixel-wise control for the sparsity of the similarity matrix.

\begin{figure}[!htb]
\begin{minipage}{0.48\linewidth}
\centerline{\includegraphics[scale=0.25]{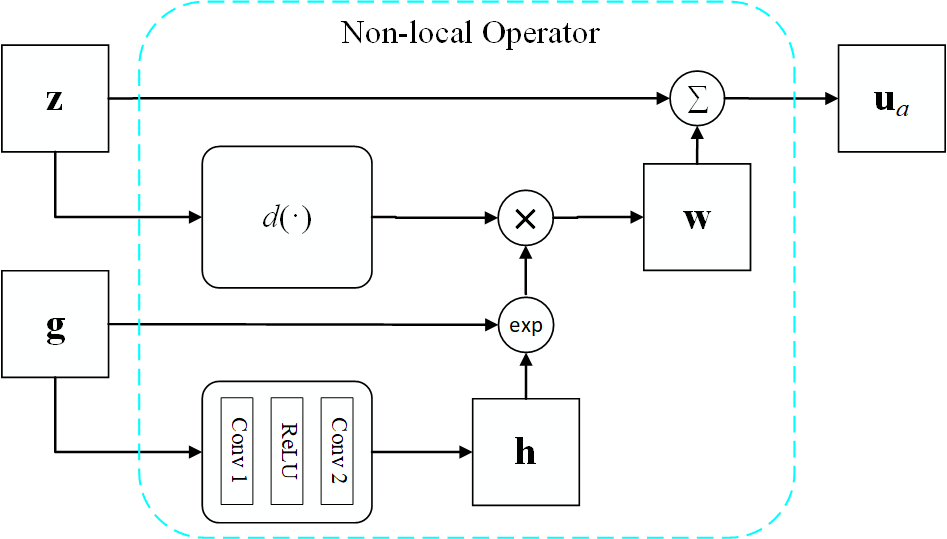}}
\centerline{(a)}
\end{minipage}
\begin{minipage}{0.48\linewidth}
\centerline{\includegraphics[scale=0.3]{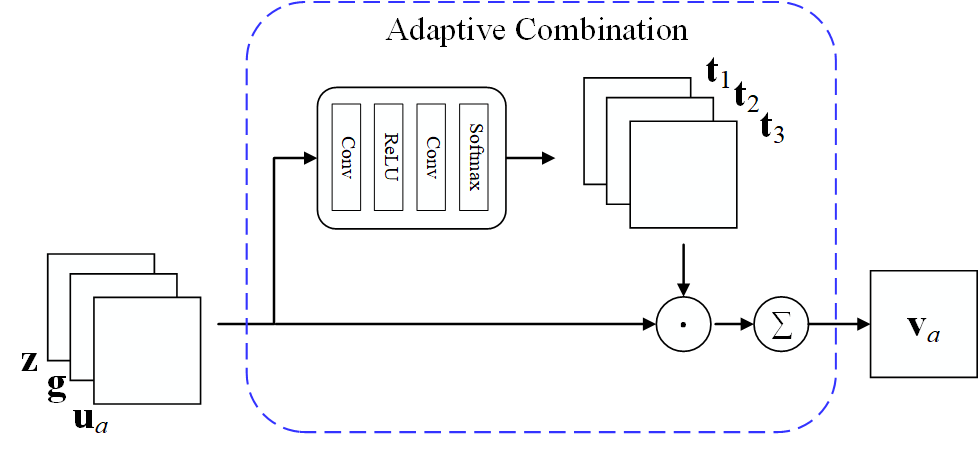}}
\centerline{(b)}
\end{minipage}
\caption{(a) Flowchart of our non-local operator.  (b) Flowchart of our adaptive combination for long skip connection.}

\label{fig:res4}
\end{figure}

The flowchart of our modified non-local operator is illustrated in Fig. \ref{fig:res4}(a). The modified non-local operators in the ARM and the REM utilize the same formulation and flowchart, but their learnable parameters are not shared during the training.
\subsubsection{Adaptive combination for long skip connection}
\label{residual}
In the architectures of the ARM and the REM, we include a long skip connection, although there may already exist multiple short or long skip connections in the backbone. Instead of delivering learned features, the purpose of this skip connection is to pass the signals from the input end to the output end. Such an input-to-output pass-by has been demonstrated to be effective and necessary in solving the SISR problem \cite{ref15}. In our architecture, after the non-local operator, three images, \textit{i.e.}, \textbf{z}, \textbf{g}, and $\mathbf{u}_{a}$, are available for the long skip connection. Their properties are different. The input \textbf{z} is the signal remained unprocessed, but it may suffer from severe artefacts. The image \textbf{g} is from the auxiliary input, which is assumed to be a clean signal. However, since the auxiliary input of one module is the output of the other one, some fake signal patterns would be introduced into \textbf{g} when we train the other module by external samples. The image $\mathbf{u}_{a}$ is from the non-local operator and much cleaner than the input \textbf{z}. Although $\mathbf{u}_{a}$ is also a processed image, all the signal patterns in $\mathbf{u}_{a}$ are from \textbf{z} itself. Therefore, all three images should be included in the long skip connection. In this work, we use a convex combination with adaptive weights to fuse them to one image. That is
\begin{equation}
\begin{aligned}
\mathbf{v}_{a}=\mathbf{t}_{1} \odot \mathbf{z}+\mathbf{t}_{2} \odot \mathbf{g} + \mathbf{t}_{3} \odot \mathbf{u}_{a} \\
\text{s.t.}     \ \mathbf{t}_{1} + \mathbf{t}_{2} +\mathbf{t}_{3}=\mathbf{1},
\end{aligned}
\end{equation}
where $\mathbf{v}_{a}$ is the output of the adaptive combination, the subscript \textit{a} indicates Module I or Module II, $\mathbf{t}_{1}$, $\mathbf{t}_{2}$, and $\mathbf{t}_{3}$ are the weight maps for \textbf{z}, \textbf{g}, and $\mathbf{u}_{a}$ respectively, $\odot$ represents the pixel-wise multiplication, and \textbf{1} denotes the map of all ones. Note that $\mathbf{t}_{1}$, $\mathbf{t}_{2}$, and $\mathbf{t}_{3}$ are three maps instead of three scalar values, which can further improve the flexibility. As demonstrated in \cite{ref67}, the identity mapping is the best option for residual learning. Thus, the constraint on the summation of $\mathbf{t}_{1}$, $\mathbf{t}_{2}$, and $\mathbf{t}_{3}$ is essential to make our long skip connection approximate the identity mapping. Obviously, the weight maps in Eq. (11) depend on the image contents and qualities of \textbf{z}, \textbf{g}, and $\mathbf{u}_{a}$. Thus, similar to the adaptive parameter \textbf{h} in Eq. (10), we employ a light network to estimate $\mathbf{t}_{1}$, $\mathbf{t}_{2}$, and $\mathbf{t}_{3}$ from \textbf{z}, \textbf{g}, and $\mathbf{u}_{a}$. This network consists of two convolutional layers, which are followed by a ReLU layer and a SoftMax operation, respectively. The SoftMax operation is performed on each pixel position to satisfy the constraint in Eq. (11). The flowchart of our adaptive combination is summarized in Fig. \ref{fig:res4}(b). In Section \ref{abla}, we also discuss the contribution of individual images, against their adaptive combination. 

\section{Experiments}
\label{exper}
In this section, we will provide implementation details and the datasets used in our experiments first. Subsequently, we compare the proposed method with state-of-the-art SR methods. Then, ablation studies are conducted to demonstrate the effectiveness of the proposed method. Finally, we apply our method to a real-world problem - the restoration of \textit{WeChat} avatar images that have undergone unknown scaling and compression.

\subsection{Implementation Details and Datasets}
\label{dataset}

The proposed method is implemented in PyTorch on a machine of NVIDIA GeForce 1080Ti. Two versions of our model are included in the following comparisons. One is named as tiny model, and the other one is called full model. The number of learnable parameters in the backbone of the tiny model is much smaller than that of the full model. Specifically, we adopt 5 residual groups from RCAN as the backbone for each module in our full model. In our tiny model, only 2 residual groups are employed for each module. Moreover, each residual group in our tiny or full model contains 12 channel attention blocks, instead of 20 attention blocks used in the original RCAN model. Thus, our full model is still much lighter than RCAN. In addition to the convolutional layers in the backbone, there are two convolutional layers in the modified non-local operator, as shown in Fig. \ref{fig:res4}(a), and two convolutional layers are used in the adaptive combination, as shown in Fig. \ref{fig:res4}(b). Among these four layers, the kernel size of the first convolutional layer in Fig. \ref{fig:res4}(a) is $3\times 3$, while the kernel sizes of the rest are $1\times 1$. The numbers of their output channel are 64, 1, 64, and 3, respectively. The maximum iteration number $J$ is empirically set to 3 for our full model and 5 for our tiny one. Correspondingly, we set $\rho_{j}$ in Eq. (8) as {0.3, 0.6, 1} for the full model and {0.2, 0.4, 0.6, 0.8, 1} for the tiny one. The parameter $\gamma$ in the loss function is simply set to 1.
 
Two widely-used compression types JPEG \cite{ref61} and WebP \cite{ref70}, are involved in our experiments. For each type of compression, we further involve 5 compression levels or quality factors (QFs). Specifically, for the JPEG compression, we have QFs of 10, 20, 30, 40, and 50. For the WebP compression, QFs of 5, 10, 20, 30, and 40 are utilized. Hence, to compress an LR image, we have a total of 10 kinds of compression configurations, viz., 2 compression types multiplied by 5 QFs. Furthermore, three scaling factors ($2\times$, $3\times$, and $4\times$) are involved in the experiments. For each up-scaling factor, we train a tiny model and a full model over all the compression configurations which leads to a total of 6 models in our experiments \textit{i.e.}, tiny model ($2\times$), full model ($2\times$), tiny model ($3\times$), full model ($3\times$), tiny model ($4\times$), full model ($4\times$). In other words, LR images with various compression configurations are mixed together to train and test these 6 models. Since our focus is on the compression of LR images rather than down-sampling kernels, the bicubic down-sampling is employed to resize images for simplicity. To prepare training and validation data, compressed LR images are generated by first down-sampling and then compressing the samples in DIV2K \cite{ref71}. Data augmentation is also performed on the training pairs by random rotation and flipping. In each training batch, we randomly crop 32 patches with the size of $48\times 48$ as LR inputs. Our models are trained by the ADAM optimizer \cite{ref69} with an initial learning rate of $10^{-4}$. The training is terminated when the performance of the model decreases on the validation set.

In addition to the training and validation data, we also require a testing dataset with ground truths to facilitate quantitative comparisons. Ground truths should be uncompressed HR images. However, most publicly available image datasets suffer from compression to some extent. As an exception, the Kodak24 dataset \cite{ref72} contains 24 lossless images, which are used to produce our testing dataset. Moreover, we capture another 76 images in various scenarios by ourselves. These images are lossless as well. For more details about these captured images, one can refer to our online supplementary materials in \cite{ref73}. In total, 100 lossless HR images are used as ground truths in our quantitative tests. By using these 100 images and the above-mentioned compression configurations, we produce 1000 testing LR images for each up-scaling factor.

Finally, to demonstrate the value of our method in real-world application, we have collected a dataset of 50 images from a social media platform. These images have undergone severe scaling and compression by unknown algorithms hidden from the users. We apply the trained models directly to these real-world images, which will be shown in Section \ref{real_app}. 

\subsection{Comparisons}
\label{compare}

\begin{table}[ht]
\caption{Quantitative Comparisons on PSNR, SSIM, IFC, and SIS.}
\setlength{\tabcolsep}{1.8mm}{        
\begin{tabular}{|c|l|c|c|c|c|}
\hline
\multirow{2}{*}{\begin{tabular}[c]{@{}c@{}}Up-scaling \\ factor\end{tabular}} & \multicolumn{1}{c|}{\multirow{2}{*}{SR Models}} & \multicolumn{4}{c|}{Criteria}                                        \\ \cline{3-6} 
                                                                              & \multicolumn{1}{c|}{}                           & PSNR           & SSIM            & IFC             & SIS             \\ \hline
\multirow{9}{*}{2}                                                            & ICSD                                            & 28.86          & 0.7903          & 1.7968          & 0.6662          \\
                                                                              & CISRDCNN                                        & 29.31          & 0.8019          & 1.8964          & 0.7094          \\
                                                                              & DnCNN + SAN                                     & 29.48          & 0.8027          & 1.8815          & 0.7223          \\
                                                                              & DnCNN + RCAN                                    & 28.67          & 0.7927          & 1.7316          & \textit{0.7540} \\
                                                                              & DnCNN + USRNet                                  & 29.49          & 0.8028          & 1.8850          & 0.7212          \\
                                                                              & USRNet                                          & 28.59          & 0.7682          & 1.6711          & 0.6381          \\
                                                                              & USRNet + NRIBP                                  & 29.47          & 0.8020          & 1.8619          & 0.7216          \\
                                                                              & Our tiny model                                  & \textit{29.97} & \textit{0.8149} & \textit{2.0978} & 0.7495          \\
                                                                              & Our full model                                  & \textbf{30.10} & \textbf{0.8181} & \textbf{2.1521} & \textbf{0.7576} \\ \hline
\multirow{9}{*}{3}                                                            & ICSD                                            & 26.82          & 0.7259          & 1.1362          & 0.4956          \\
                                                                              & CISRDCNN                                        & 27.37          & 0.7391          & 1.2406          & 0.5287          \\
                                                                              & DnCNN + SAN                                     & 27.40          & 0.7384          & 1.2151          & 0.5374          \\
                                                                              & DnCNN + RCAN                                    & 26.75          & 0.7245          & 1.0905          & 0.5845          \\
                                                                              & DnCNN + USRNet                                  & 27.43          & 0.7392          & 1.2208          & 0.5399          \\
                                                                              & USRNet                                          & 26.86          & 0.7159          & 1.1387          & 0.4637          \\
                                                                              & USRNet + NRIBP                                  & 27.40          & 0.7378          & 1.2022          & 0.5397          \\
                                                                              & Our tiny model                                  & \textit{27.84} & \textit{0.7536} & \textit{1.3947} & \textit{0.5864} \\
                                                                              & Our full model                                  & \textbf{27.94} & \textbf{0.7564} & \textbf{1.4355} & \textbf{0.5893} \\ \hline
\multirow{9}{*}{4}                                                            & ICSD                                            & 25.54          & 0.6820          & 0.7974          & 0.3702          \\
                                                                              & CISRDCNN                                        & 25.99          & 0.6959          & 0.8995          & 0.3691          \\
                                                                              & DnCNN + SAN                                     & 26.16          & 0.6954          & 0.8767          & 0.3843          \\
                                                                              & DnCNN + RCAN                                    & 25.65          & 0.6808          & 0.7799          & 0.4167          \\
                                                                              & DnCNN + USRNet                                  & 26.18          & 0.6964          & 0.8811          & 0.3863          \\
                                                                              & USRNet                                          & 25.65          & 0.6759          & 0.8175          & 0.3251          \\
                                                                              & USRNet + NRIBP                                  & 26.16          & 0.6950          & 0.8670          & 0.3883          \\
                                                                              & Our tiny model                                  & \textit{26.55} & \textit{0.7117} & \textit{1.0267} & \textit{0.4404} \\
                                                                              & Our full model                                  & \textbf{26.62} & \textbf{0.7138} & \textbf{1.0504} & \textbf{0.4418} \\ \hline
\end{tabular}}
\label{tab1}
\end{table}

\begin{figure*}[]

\begin{minipage}{1\linewidth}
\centerline{\includegraphics[height=17cm,width=17cm]{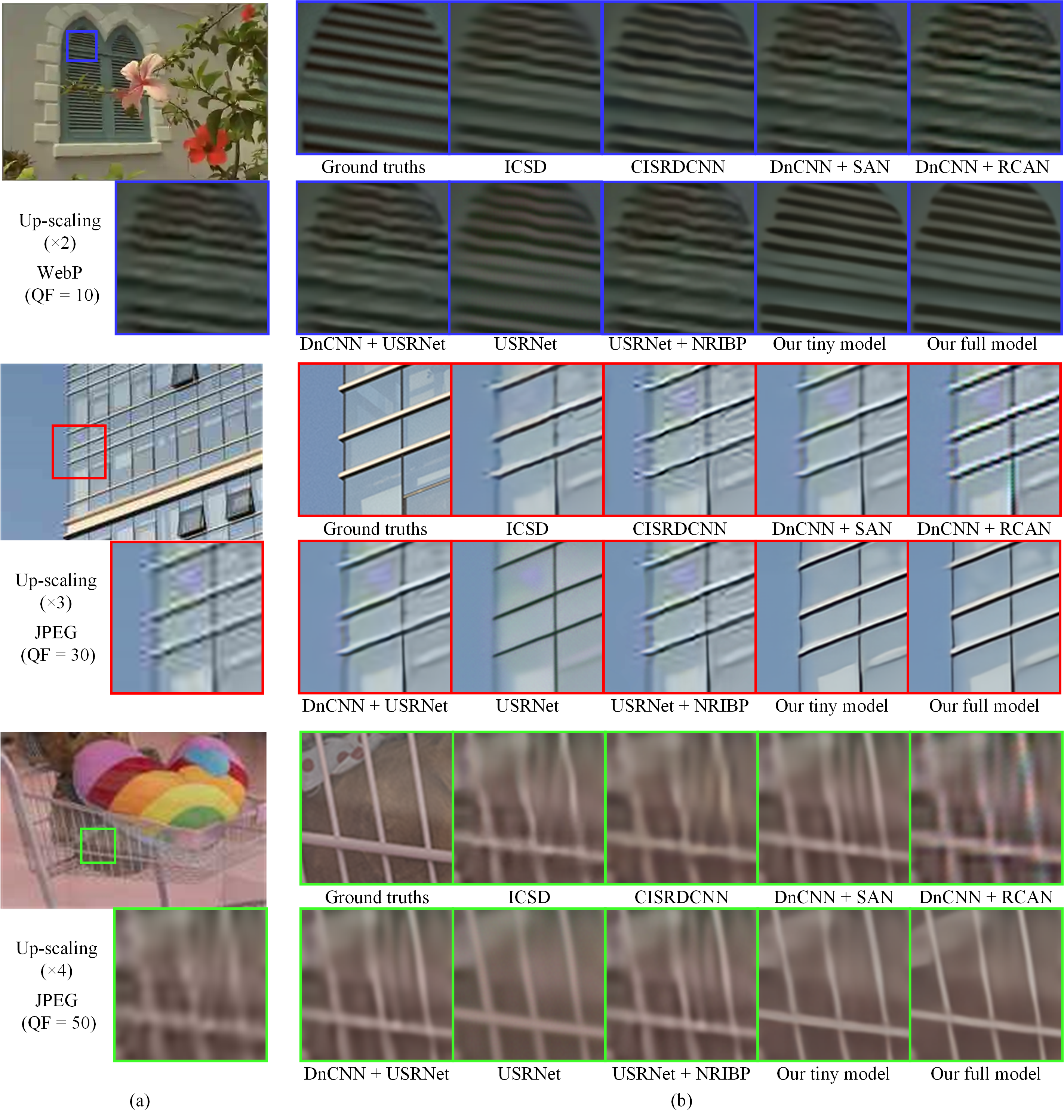}}
\end{minipage}

\caption{Visual comparisons. (a) Several compressed LR images for testing. (b) Ground truths and results of different SISR methods.}
\label{fig:res5}
\end{figure*}

To show the effectiveness of the proposed method, we compare it with some state-of-the-art SISR models, including ICSD \cite{ref34}, CISRDCNN \cite{ref35}, SAN \cite{ref20}, RCAN \cite{ref19}, and USRNet \cite{ref43}. Among these competitors, ICSD is a traditional method, while all the others are based on DCNN. The models of ICSD and CISRDCNN are specifically designed for CISR. The models of SAN and RCAN are proposed for clean LR images. The USRNet model is applicable to either clean images or noisy and blurred images. For fair comparisons, we add a pre-processing before SAN, RCAN, and USRNet to reduce compression artefacts. Here, the pre-processing is achieved by DnCNN \cite{ref53}, which is a widely-used model trained to reduce compression artefacts for a wide range of QFs. The USRNet model can further incorporate noise levels and blur kernels of LR images. It would be interesting to see whether the performance of SISR is satisfying when compression artefacts are treated as noises and blurring. Thus, we also provide the SR results that are obtained by applying USRNet alone. To estimate noise levels and blur kernels for USRNet, the methods in \cite{ref74} and \cite{ref75} are adopted in our experiments, respectively. Moreover, the method of NRIBP \cite{ref40} can be combined with SR models to suppress the noises and artefacts in SR results, \textit{e.g.}, USRNet + NRIBP. The codes of ICSD, CISRDCNN and NRIBP are implemented by ourselves, while the rest are provided by their authors. Here, we provide visual and quantitative comparisons in Fig. \ref{fig:res5} and Table \ref{tab1}. One can refer to our online materials in \cite{ref73} for more results in comparison with more SISR methods, including A+ \cite{ref8} and EDSR \cite{ref16}.

In Fig. \ref{fig:res5}(a), we show three examples of testing LR images, which are heavily compressed by JPEG or WebP. In each image, we highlight an image region, whose results from different SISR models are exhibited in Fig. \ref{fig:res5}(b). From the results of highlighted regions, we can see that results from some competitors suffer from conspicuous artefacts. At the same time, the results from other compared SISR models are over-smoothed. In contrast, our models can successfully retrieve sharp edges as well as remove artefacts. Moreover, the results from our tiny model are visually comparable with those from our full model, although the former is much lighter. 

\begin{figure*}[ht]
\centerline{\includegraphics[scale=0.6]{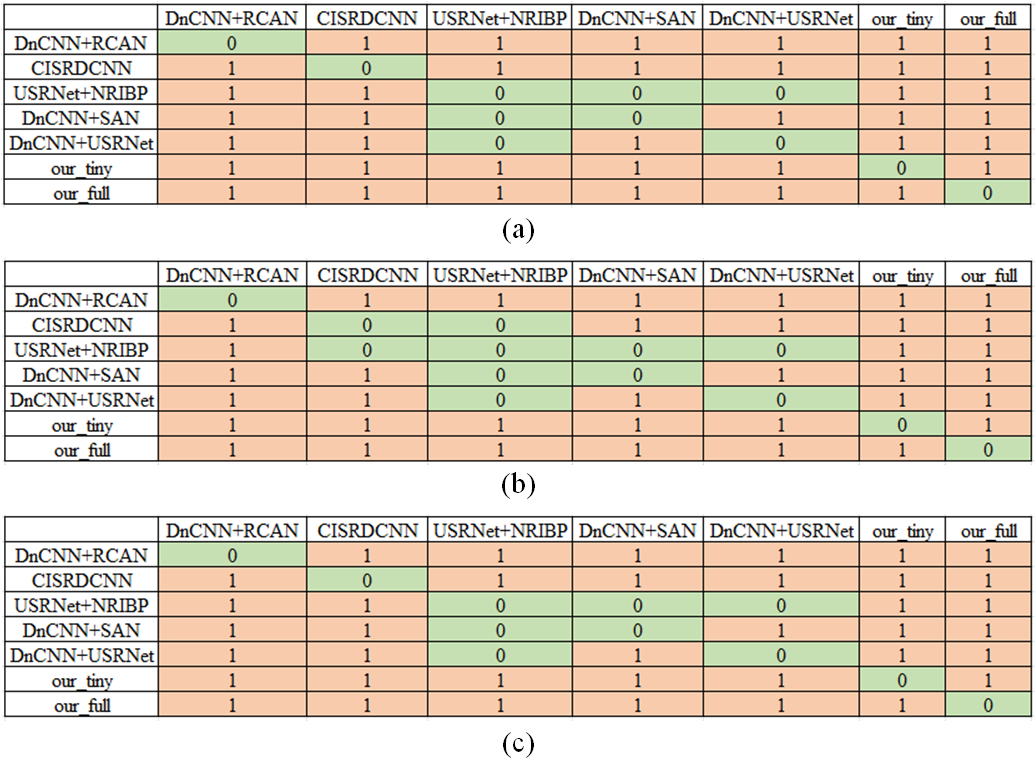}}
\caption{Statistical significance testing on PSNR in Table \ref{tab1} for three scale factors. (a) $2\times$ (b) $3\times$ (c) $4\times$.}
\label{fig:sign1}
\end{figure*} 

Four criteria are used to quantitatively measure the performance of different SR methods. They are peak signal to noise ratio (PSNR), structural similarity (SSIM) index \cite{ref76}, information fidelity criterion (IFC) \cite{ref77}, and structure-texture decomposition for image quality assessment of SRIs (SIS) \cite{ref78}. PSNR and SSIM are widely adopted in the evaluation of SRIs. And it has been demonstrated in \cite{ref1} and \cite{ref78} that IFC and SIS have relatively high correlations with the perceptual quality of SRIs. Therefore, it is appropriate to include these four criteria in quantitative comparisons. For all these criteria, larger values imply better performance. Quantitative results of competitors and our models are provided in Table \ref{tab1}, which contains three up-scaling factors. For each up-scaling factor, the listed values are the average results over 1000 testing images, \textit{i.e.}, 100 images multiplied by 10 compression configurations. The best performance is highlighted in bold, and the second-best results are distinguished by italics. From Table \ref{tab1}, we can see that our full model achieves an improvement of 0.5-0.6 dB on PSNR. On the other three criteria, its superiority to the competitors is also significant. Notably, even our tiny model can achieve very good performance on all the criteria, although it is much lighter than our full model.

To further demonstrate the significant improvement of our models, we conduct the statistical testing in Table \ref{tab1}, checks whether our method is statistically distinguishable from the competitors. Specifically, the paired-samples T-test is conducted on each up-sampling scale. In each scale, there are 1000 samples which approximately follows normal distributions. Due to the limitation space, we only perform the statistical significance test on PSNR. The results of this statistical significance test are shown in Fig. \ref{fig:sign1}, where the array element of “1” presents the p-value is less than 0.05, implying that there is significant difference of performance between two SR models. Otherwise, the array element is filled with “0”. From Fig. \ref{fig:sign1}, we can find that the performance differences between our model (either the tiny one or the full one) and all the competitors are statistically significant. Besides, as shown in Table \ref{tab1}, our model achieves the highest PSNR. Thus, we can conclude that our model is significantly better than the compared ones.

To show the performance of the proposed models across different compression configurations, we list detailed results based on PSNR in Table \ref{tab2}. There are 100 testing images for each compression configuration. Thus, each PSNR value in Table \ref{tab2} are the average result over 100 images. Detailed results based on SSIM, IFC, and SIS can be found in \cite{ref73}. In conclusion, the visual and quantitative comparisons in Fig. \ref{fig:res5} and Tables \ref{tab1}-\ref{tab2} demonstrate the effectiveness of our models.

\begin{table}[ht]
\centering
\caption{Quantitative Comparisons for Different Compression Configurations Based on PSNR.}
\renewcommand\arraystretch{0.8}
\centerline{
\setlength{\tabcolsep}{1mm}{        
\begin{tabular}{|c|l|c|c|c|c|c|c|c|c|c|c|}
\hline
\multirow{2}{*}{\begin{tabular}[c]{@{}c@{}}Up-scaling\\  factor\end{tabular}} & \multicolumn{1}{c|}{\multirow{2}{*}{SR Models}} & \multicolumn{10}{c|}{Compression Configurations (Compression type \& QF)}                                                                                                                                                                                                                                                                                                                                                                                                                                                                                                    \\ \cline{3-12} 
                                                                              & \multicolumn{1}{c|}{}                           & \begin{tabular}[c]{@{}c@{}}JPEG \\ \& 10\end{tabular} & \begin{tabular}[c]{@{}c@{}}JPEG \\ \& 20\end{tabular} & \begin{tabular}[c]{@{}c@{}}JPEG \\ \& 30\end{tabular} & \begin{tabular}[c]{@{}c@{}}JPEG \\ \& 40\end{tabular} & \begin{tabular}[c]{@{}c@{}}JPEG \\ \& 50\end{tabular} & \begin{tabular}[c]{@{}c@{}}WebP \\ \& 5\end{tabular} & \begin{tabular}[c]{@{}c@{}}WebP \\ \& 10\end{tabular} & \begin{tabular}[c]{@{}c@{}}WebP \\ \& 20\end{tabular} & \begin{tabular}[c]{@{}c@{}}WebP \\ \& 30\end{tabular} & \begin{tabular}[c]{@{}c@{}}WebP \\ \& 40\end{tabular} \\ \hline
\multirow{9}{*}{2}                                                            & ICSD                                            & 26.96                                                 & 28.31                                                 & 29.02                                                 & 29.52                                                 & 29.90                                                  & 27.47                                                & 28.19                                                  & 29.11                                                 & 29.76                                                 & 30.30                                                 \\
                                                                              & CISRDCNN                                        & 27.51                                                 & 28.91                                                 & 29.62                                                 & 30.07                                                 & 30.41                                                 & 27.88                                                & 28.57                                                 & 29.45                                                 & 30.07                                                 & 30.54                                                 \\
                                                                              & DnCNN + SAN                                     & 27.68                                                 & 29.14                                                 & 29.91                                                 & 30.41                                                 & 30.82                                                 & 27.72                                                & 28.46                                                 & 29.48                                                 & 30.27                                                 & 30.92                                                 \\
                                                                              & DnCNN + RCAN                                    & 27.35                                                 & 28.49                                                 & 29.04                                                 & 29.38                                                 & 29.63                                                 & 27.36                                                & 27.93                                                 & 28.67                                                 & 29.21                                                 & 29.64                                                 \\
                                                                              & DnCNN + USRNet                                  & 27.68                                                 & 29.12                                                 & 29.90                                                  & 30.42                                                 & 30.82                                                 & 27.74                                                & 28.48                                                 & 29.51                                                 & 30.28                                                 & 30.93                                                 \\
                                                                              & USRNet                                          & 27.23                                                 & 28.41                                                 & 28.93                                                 & 29.20                                                  & 29.38                                                 & 27.60                                                 & 28.09                                                 & 28.69                                                 & 29.05                                                 & 29.28                                                 \\
                                                                              & USRNet + NRIBP                                  & 27.57                                                 & 29.05                                                 & 29.87                                                 & 30.40                                                  & 30.81                                                 & 27.73                                                & 28.47                                                 & 29.52                                                 & 30.30                                                  & 30.96                                                 \\
                                                                              & Our tiny model                                  & \textit{28.01}                                        & \textit{29.51}                                        & \textit{30.31}                                        & \textit{30.84}                                        & \textit{31.25}                                        & \textit{28.41}                                       & \textit{29.11}                                        & \textit{30.07}                                        & \textit{30.81}                                        & \textit{31.41}                                        \\
                                                                              & Our full model                                  & \textbf{28.11}                                        & \textbf{29.61}                                        & \textbf{30.43}                                        & \textbf{30.95}                                        & \textbf{31.37}                                        & \textbf{28.56}                                       & \textbf{29.26}                                        & \textbf{30.20}                                         & \textbf{30.93}                                        & \textbf{31.53}                                        \\ \hline
\multirow{9}{*}{3}                                                            & ICSD                                            & 25.35                                                 & 26.41                                                 & 26.94                                                 & 27.31                                                 & 27.58                                                 & 25.82                                                & 26.41                                                 & 27.02                                                 & 27.49                                                 & 27.88                                                 \\
                                                                              & CISRDCNN                                        & 25.90                                                  & 27.05                                                 & 27.62                                                 & 27.98                                                 & 28.22                                                 & 26.27                                                & 26.84                                                 & 27.53                                                 & 28.00                                                    & 28.36                                                 \\
                                                                              & DnCNN + SAN                                     & 25.97                                                 & 27.09                                                 & 27.67                                                 & 28.06                                                 & 28.34                                                 & 26.09                                                & 26.69                                                 & 27.49                                                 & 28.08                                                 & 28.55                                                 \\
                                                                              & DnCNN + RCAN                                    & 25.70                                                  & 26.59                                                 & 27.02                                                 & 27.26                                                 & 27.45                                                 & 25.74                                                & 26.20                                                  & 26.78                                                 & 27.19                                                 & 27.51                                                 \\
                                                                              & DnCNN + USRNet                                  & 25.96                                                 & 27.09                                                 & 27.68                                                 & 28.08                                                 & 28.37                                                 & 26.09                                                & 26.72                                                 & 27.53                                                 & 28.13                                                 & 28.61                                                 \\
                                                                              & USRNet                                          & 25.59                                                 & 26.63                                                 & 27.09                                                 & 27.35                                                 & 27.52                                                 & 26.03                                                & 26.48                                                 & 27.02                                                 & 27.33                                                 & 27.55                                                 \\
                                                                              & USRNet + NRIBP                                  & 25.87                                                 & 27.03                                                 & 27.65                                                 & 28.05                                                 & 28.36                                                 & 26.08                                                & 26.70                                                  & 27.53                                                 & 28.13                                                 & 28.62                                                 \\
                                                                              & Our tiny model                                  & \textit{26.22}                                        & \textit{27.43}                                        & \textit{28.08}                                        & \textit{28.49}                                        & \textit{28.78}                                        & \textit{26.65}                                       & \textit{27.23}                                        & \textit{27.98}                                        & \textit{28.54}                                        & \textit{28.99}                                        \\
                                                                              & Our full model                                  & \textbf{26.31}                                        & \textbf{27.53}                                        & \textbf{28.18}                                        & \textbf{28.59}                                        & \textbf{28.88}                                        & \textbf{26.78}                                       & \textbf{27.35}                                        & \textbf{28.09}                                        & \textbf{28.63}                                        & \textbf{29.07}                                        \\ \hline
\multirow{9}{*}{4}                                                            & ICSD                                            & 24.29                                                 & 25.20                                                  & 25.65                                                 & 25.97                                                 & 26.19                                                 & 24.75                                                & 25.12                                                 & 25.73                                                 & 26.11                                                 & 26.45                                                 \\
                                                                              & CISRDCNN                                        & 24.73                                                 & 25.70                                                  & 26.17                                                 & 26.46                                                 & 26.66                                                 & 25.09                                                & 25.56                                                 & 26.14                                                 & 26.53                                                 & 26.81                                                 \\
                                                                              & DnCNN + SAN                                     & 24.88                                                 & 25.87                                                 & 26.37                                                 & 26.70                                                  & 26.94                                                 & 25.04                                                & 25.56                                                 & 26.27                                                 & 26.78                                                 & 27.17                                                 \\
                                                                              & DnCNN + RCAN                                    & 24.68                                                 & 25.50                                                  & 25.87                                                 & 26.10                                                  & 26.27                                                 & 24.75                                                & 25.17                                                 & 25.71                                                 & 26.06                                                 & 26.34                                                 \\
                                                                              & DnCNN + USRNet                                  & 24.87                                                 & 25.87                                                 & 26.39                                                 & 26.72                                                 & 26.97                                                 & 25.05                                                & 25.59                                                 & 26.31                                                 & 26.83                                                 & 27.23                                                 \\
                                                                              & USRNet                                          & 24.48                                                 & 25.40                                                  & 25.81                                                 & 26.05                                                 & 26.21                                                 & 24.95                                                & 25.35                                                 & 25.82                                                 & 26.09                                                 & 26.28                                                 \\
                                                                              & USRNet + NRIBP                                  & 24.81                                                 & 25.83                                                 & 26.36                                                 & 26.70                                                  & 26.96                                                 & 25.04                                                & 25.58                                                 & 26.30                                                  & 26.83                                                 & 27.23                                                 \\
                                                                              & Our tiny model                                  & \textit{25.12}                                        & \textit{26.21}                                        & \textit{26.76}                                        & \textit{27.11}                                        & \textit{27.36}                                        & \textit{25.53}                                       & \textit{26.03}                                        & \textit{26.71}                                        & \textit{27.19}                                        & \textit{27.56}                                        \\
                                                                              & Our full model                                  & \textbf{25.17}                                        & \textbf{26.25}                                        & \textbf{26.81}                                        & \textbf{27.17}                                        & \textbf{27.44}                                        & \textbf{25.62}                                       & \textbf{26.13}                                        & \textbf{26.78}                                        & \textbf{27.25}                                        & \textbf{27.62}                                        \\ \hline
\end{tabular}}}
\label{tab2}
\end{table}

In addition to SISR results, we would like to compare our results of compression artefacts removal, \textit{i.e.}, the output of Module I, with several classical and state-of-the-art methods, including SA-DCT \cite{ref49}, TNRD \cite{ref57}, DnCNN \cite{ref53}, and DCSC \cite{ref55}. Quantitative and visual comparisons can be accessed in our online materials \cite{ref73}. It can be seen from these results that our technique has excellent performance in reducing artefacts and recovering details.  

Moreover, we perform model parameters and runtime comparisons with other models. The comparisons are shown in Table \ref{runtime}, where the runtime is calculated on an RGB image with the size of $3\times128\times128$. All of the models are run on the device we mentioned in Section \ref{dataset}. From the comparison, it can be seen that our models, both the tiny and the full model, use much less parameters to achieve the best performance due to the iteration optimization. Obviously, the iteration optimization procedure costs more running time, and the reason, that the runtime of our tiny model is longer than the one of our full model, is five iteration step for the tiny model while three steps for the full model.

\begin{table}[]
\centering
\caption{Comparison of parameters and runtime.}
\label{runtime}
\begin{tabular}{|c|c|c|}
\hline
               & params(M) & runtime(ms) \\ \hline
CISRDCNN       & 1.28      & 18          \\ \hline
DnCNN+SAN      & 16.38     & 287         \\ \hline
DnCNN+RCAN     & 16.08     & 66          \\ \hline
DnCNN+USRNet   & 17.69     & 56          \\ \hline
Our tiny model & 3.97      & 211         \\ \hline
Our full model & 9.56      & 179         \\ \hline
\end{tabular}
\end{table}

\subsection{Ablation Studies}

\subsubsection{Outputs in different iterations}
\label{iter_output}
\begin{figure*}[ht]
\centerline{\includegraphics[scale=0.8]{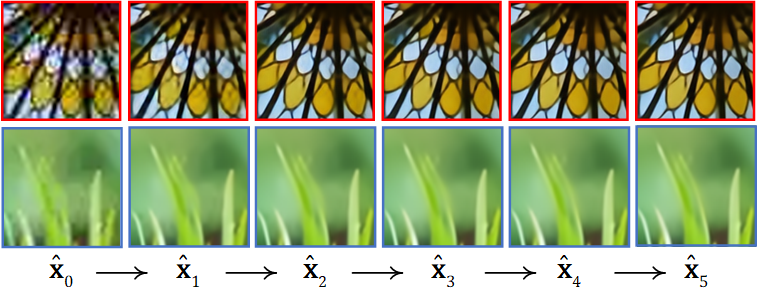}}
\caption{HR estimations in different iterations of the proposed framework.}
\label{fig:res6}
\end{figure*} 
Since the proposed framework produces its results in a recursive manner, it is interesting to investigate the estimated HR output $\hat{\mathbf{x}}$ in different iterations. Two examples are shown in Fig. \ref{fig:res6}, which are the outputs of our tiny model for LR images compressed by JPEG with QF = 10 and the scaling factor is 2. From the visual results, we can see that the initial estimation $\hat{\mathbf{x}}_{0}$ is very rough. The quality of the estimation gets better with each iteration. After a few iterations, the outputs of each iteration become indistinguishable, indicating the recursive algorithm has converged.

\subsubsection{Comparison of parallel model, series model, and our framework}
\label{compare2}
\begin{figure*}[ht]
\centerline{\includegraphics[scale=0.2]{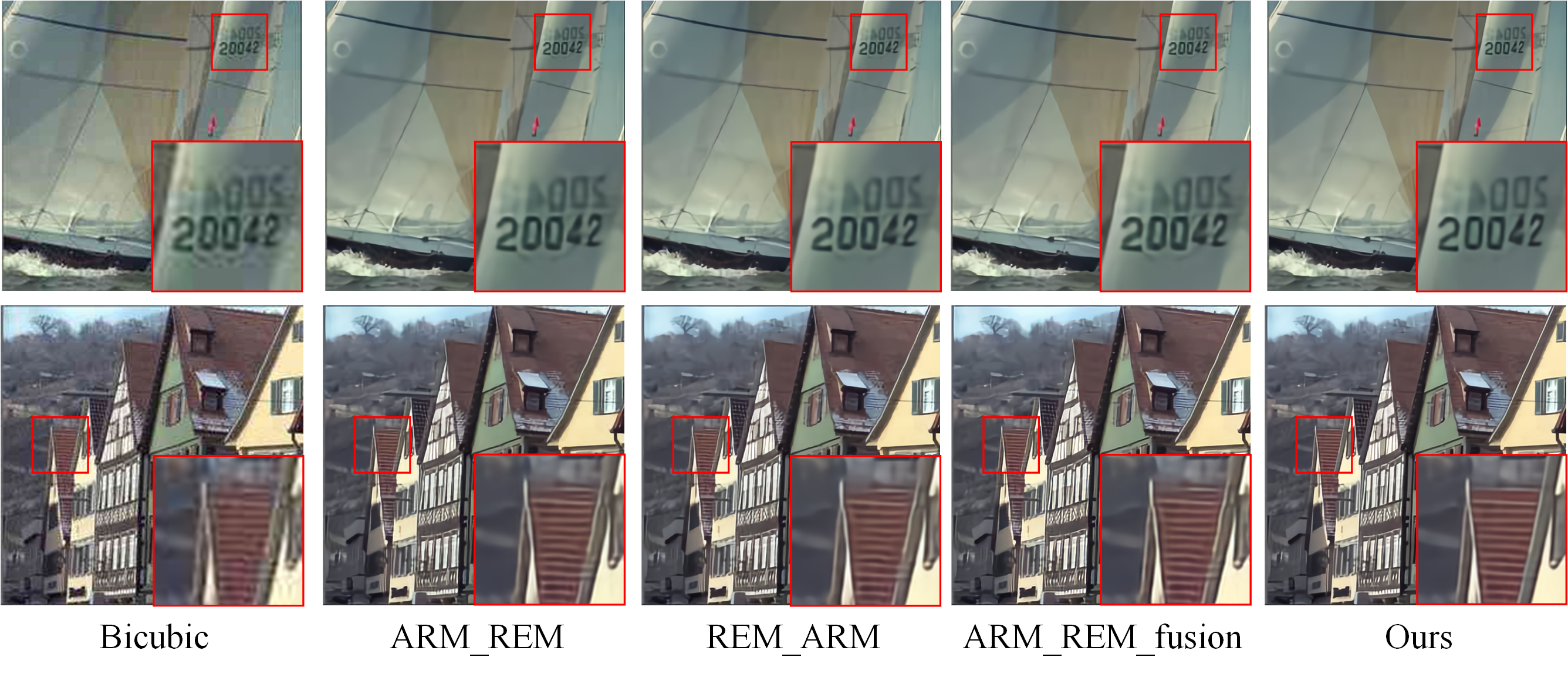}}
\caption{Visual comparisons of two images with JPEG compression QF = 30.}
\label{fig:res7}
\end{figure*}

In order to demonstrate the effectiveness of the proposed parallel and series integration model, we make a comparison with a parallel model and two series models. To make a fair comparison, the architectures of the two modules in all the parallel and series models are the same as our tiny model. For the parallel model, we bicubicly up-sample the output of the ARM, then fuse it with the output of the REM by a convolutional layer to reconstruct an RGB output. The kernel size of this convolutional layer is $3\times 3$. There are two kinds of series models, \textit{i.e.}, the ARM followed by the REM and the REM followed by the ARM. For the sake of simplicity, all models are only trained with LR images compressed by JPEG (QF = 30) and its corresponding $2\times$ HR images. The quantitative results are reported in Table. \ref{tab4}, and visual results are shown in Fig. \ref{fig:res7}. ``ARM\underline{~~}REM'' represents the series model that the ARM is followed by the REM while ``REM\underline{~~}ARM'' represents the transposition of the two modules. ``ARM\underline{~~}REM\underline{~~}fusion'' represents the parallel model. As we can see that, the proposed parallel and series integration model outperforms other two kinds of models in both quantitative and visual comparisons.

\begin{table}[ht]
\centering
\caption{Average PSNR(dB) Results of a Parallel Model, two Series Models and Our Model.}
\label{tab4}
\begin{tabular}{|l|l|l|l|l|}
\hline
     & ARM\underline{~~}REM                   & REM\underline{~~}ARM                   & ARM\underline{~~}REM\underline{~~}fusion           & Our tiny model                      \\ \hline
PNSR & \multicolumn{1}{c|}{30.04} & \multicolumn{1}{c|}{30.00} & \multicolumn{1}{c|}{29.95} & \multicolumn{1}{c|}{30.34} \\ \hline
\end{tabular}
\end{table}

\begin{table}[]
\centering
\caption{Comparison between models that two modules are shared or not.} 
\label{shared}
\begin{tabular}{|c|c|c|c|}
\hline
                 & PSNR  & SSIM   & params(M) \\ \hline
Shared (Ours)   & 27.38 & 0.7787 & 9.56       \\ \hline
Not shared (Variant)  & 27.39 & 0.7790 & 28.68      \\ \hline
\end{tabular}
\end{table}

\begin{figure}[ht]
\centerline{\includegraphics[scale=0.7]{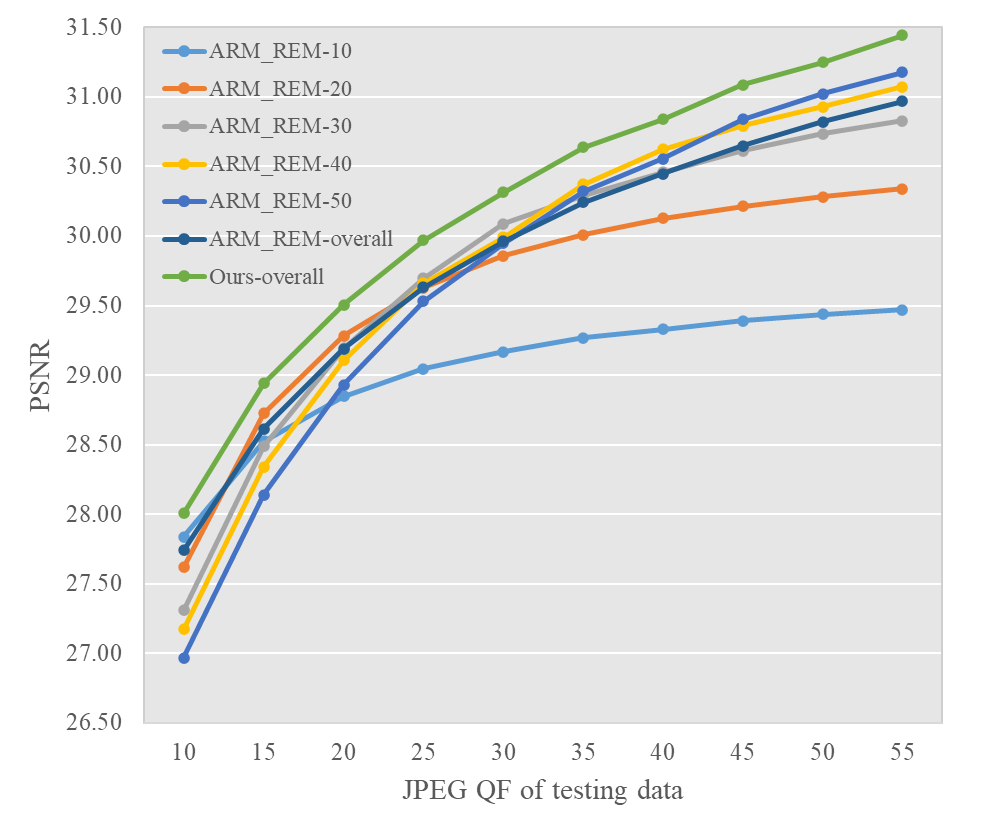}}
\caption{Average PSNR(dB) curves of different models on different compression QFs. Only scale factor of 2 are considered in this experiments.}
\label{fig:res8}
\end{figure}

\subsubsection{Shared weights vs Not Shared weights}
From the formulation in Section \ref{overall}, our model only consists of two modules, \textit{i.e.}, ARM and REM, in the recursive optimization. Therefore, the parameters of ARM are shared in each iteration, so does REM. And our models achieve the best performance as shown in Section \ref{compare}. Moreover, we conduct an ablation study to investigate the problem of parameter sharing. Specifically, our full model with scale factor 2 is compared with its variant. The variant has exactly the same architecture with our full model but not share parameters in different iterations, and thus resulting in 3 times more parameters. This study is conducted on all the ten compression configurations, and average PSNR and SSIM of these configurations on the DIV2K validation set are recorded in Table \ref{shared}. Besides, we also include the model parameter numbers in this table. From Table \ref{shared}, we see that, the variant (not shared parameters) can only achieve slight performance improvement at the expense of large number of parameters. These results demonstrate sharing parameters among iterations is beneficial in the proposed method.

\subsubsection{Analysis for different compression configurations}
\label{diff_compress}
A benefit of the proposed model is its capacity of handling different compression QFs. The recursive optimization with feedback helps to reduce the dependency on specific compression QF. To demonstrate the benefit of our model, we compare the performance of our model and 6 series models on the testing data with different compression QFs. The reasons of selecting series models as our competitors are that the series models are more common than parallel models and the series models beat the parallel model in Section \ref{compare2}. The configurations of these models are the same as the ones mentioned in Section \ref{compare2}. Fig. \ref{fig:res8} shows the performance curves of different models. Specifically, the training data with 5 JPEG compression QFs (QF=10,20,30,40,50) is adopted to train our tiny model which is represented by ``Ours-overall'' as well as a series model called ``ARM\underline{~~}REM-overall''. And we consider the 5 series models trained on the data with a single compression QF, \textit{e.g.}, ``ARM\underline{~~}REM-10'' represents the series model trained on the data with fixed compression QF=10. The compression QFs of the testing data spread from 10 to 55. Some of these QFs are not included in the QFs of the training data, \textit{e.g.}, QF=15, 25, 35, etc. From the performance curves, we have the following observations. First, the series model trained on the data with a specific QF achieves the best performance against other series models when the compression QF of the testing data matches the one of its training data. To be specific, ``ARM\underline{~~}REM-10'' beats other series models when the testing data are compressed by JPEG (QF = 10) and so are the others. However, the performances of these models deteriorate when the compression QF of the testing data mismatches the one of the training data. Second, the series model ``ARM\underline{~~}REM-overall'' achieves comparable performance on the testing data with every compression QF, whereas it fails to defeat other series models on the testing data with their corresponding compression QF. The above observations imply that pursuing the best performance means sacrificing generalization for the series models. And our model outperforms other series models on the testing data with every compression QF and achieves the best performance without sacrificing generalization. Even though the compression QF of the testing data mismatches the one of the training data, our model still attain the best performance, e.g, QF=15, 25, 35, etc. We attribute our model’s good generalization performances to the feedback routes between the ARM and the REM.

\subsubsection{Effectiveness of the modified non-local operator and the adaptive combination for residual learning}
\label{abla}
In this sub-section, we would like to conduct ablation studies to show the effectiveness of the modified non-local operator and the adaptive combination in our model. For simplicity, only our tiny model is investigated here, and only PSNR values are recorded. Moreover, this experiments are only performed on the case that LR images compressed by JPEG with QF = 10 are super-resolved by a factor of 2.

 \begin{table}[ht]
\centering
\caption{Ablation Studies of the Modified Non-local Operator and the Adaptive Combination Based on PNSR.}
\label{tab5}
\begin{threeparttable}  

\begin{tabular}{|c|l|c|c|l|}
\hline
\multicolumn{2}{|c|}{w/o\tnote{\dag}  non-local operator} & \multicolumn{2}{c|}{traditional non-local operator} & \multicolumn{1}{c|}{Ours} \\ \hline
\multicolumn{2}{|c|}{27.94}                  & \multicolumn{2}{c|}{27.95}                          & \multirow{3}{*}{28.01}    \\ \cline{1-4}
w/o residual   & z as residual               & g as residual            & $\mathbf{u}_{a}$ as residual           &                           \\ \cline{1-4}
27.90          & \multicolumn{1}{c|}{27.95}  & 27.96                    & 27.97                    &                           \\ \hline
\end{tabular}
\begin{tablenotes}
        \footnotesize
        \item[\dag] "w/o" means without.
      \end{tablenotes}
  \end{threeparttable}
\end{table}

\begin{figure}[!ht]
\begin{minipage}{1\linewidth}
\centerline{\includegraphics[scale=0.5]{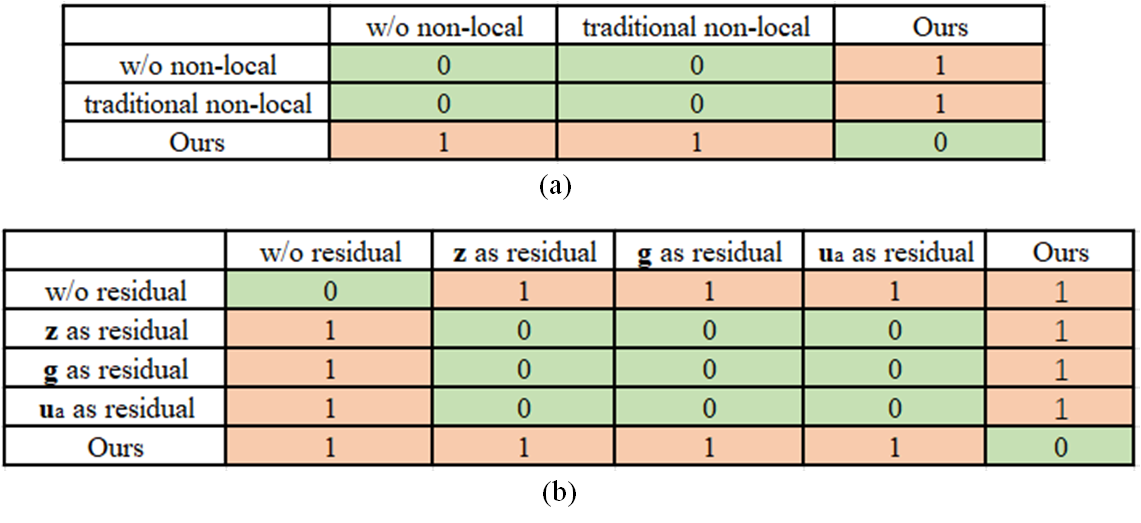}}
\end{minipage}
\centering
\caption{Statistical significance testing on PSNR in Table \ref{tab5} for two ablation studies. (a) Test for different configurations of non-local operators. (b) Test for different configurations of adaptive residual learning.}
\label{fig:ss}
\end{figure} 

We investigate the impact of the non-local operator by removing it and replacing it with a traditional non-local operator. For the traditional non-local operator, we empirically set all the elements in \textbf{h} to 30. Similarly, we can investigate the effect of the adaptive combination for skip connection by using only one of \textbf{z}, \textbf{g}, and $\mathbf{u}_{a}$ for the global residual learning. Besides, the result from the model without such a long skip connection is also recorded. All these results are provided in Table \ref{tab5}, which demonstrate the non-local operator and the adaptive combination are beneficial. Similar to Section \ref{compare}, we further measure the statistical significance for the results in Table \ref{tab5}, \textit{i.e.}, checking the significance of our modified non-local operator and the proposed adaptive residual learning. There are 100 samples for each configuration, and the results are provided in Fig. \ref{fig:ss}. From Fig. \ref{fig:ss}(a), we can see that the performance differences between the model with our non-local operator and the one with other two configurations are significant. And the results in Fig. \ref{fig:ss}(b) demonstrate that the adaptive combination of three images achieve significantly different performance compared with individual skip connections. By considering the results in Table \ref{tab5} and Fig. \ref{fig:ss}, it can be concluded that the modified non-local operator and the adaptive combination for residual learning are nontrivial.

\begin{figure*}[]

\begin{minipage}{1\linewidth}
\centerline{\includegraphics[height=16cm,width=16cm]{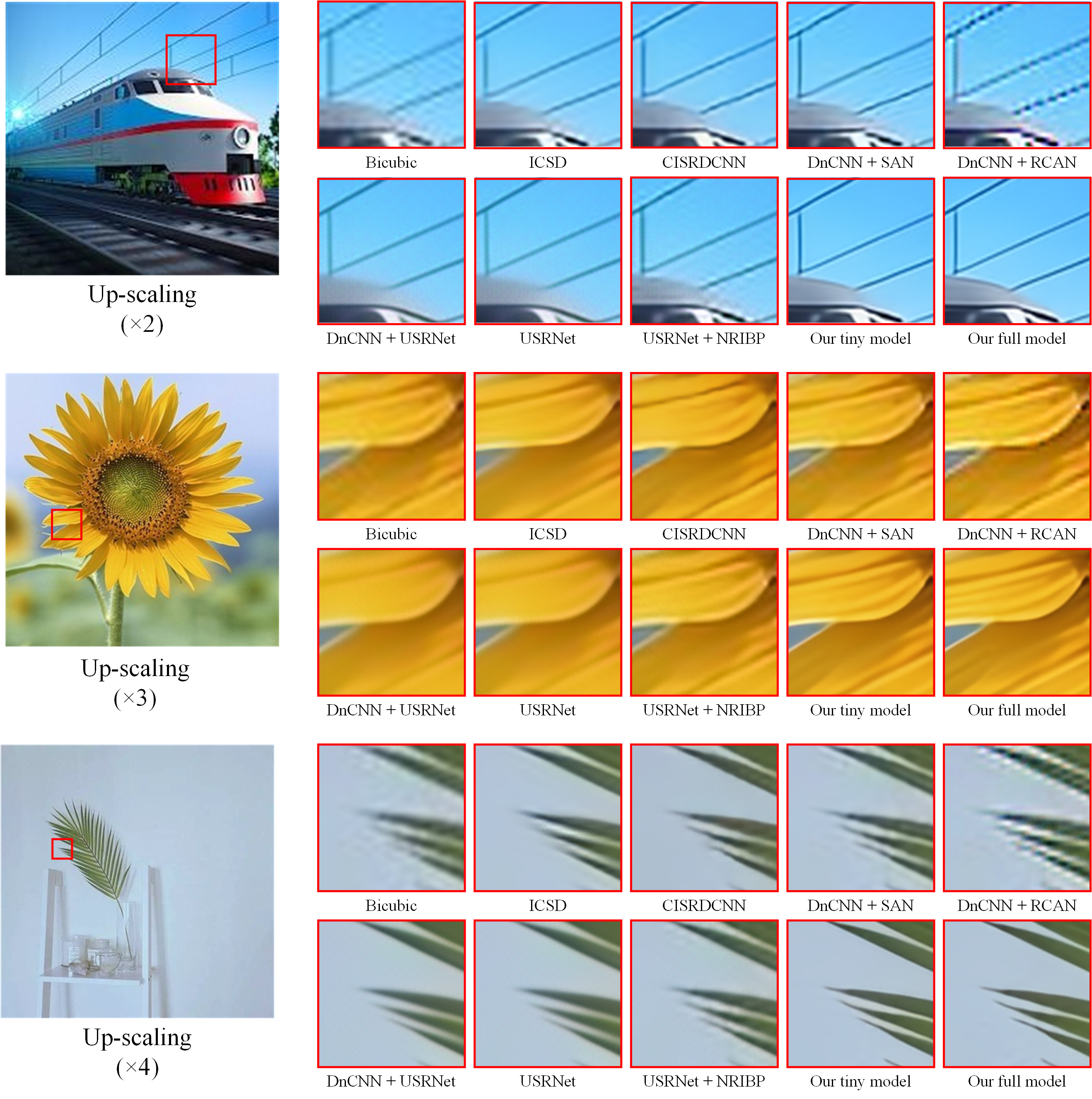}}
\end{minipage}
\centering
\caption{Visual comparisons on the collected avatar images dataset.}
\label{fig:real}
\end{figure*} 

\begin{table}[t]
\centering
\caption{Quantitative Comparisons Based on CaHDC.}
\setlength{\tabcolsep}{5.6mm}{
\begin{tabular}{|l|c|c|c|}
\hline
\multicolumn{1}{|c|}{\multirow{2}{*}{SR Models}} & \multicolumn{3}{c|}{Up-scaling factor}           \\ \cline{2-4} 
\multicolumn{1}{|c|}{}                           & 2              & 3              & 4              \\ \hline
ICSD                                             & 43.20           & 38.03          & 31.89          \\ \hline
CISRDCNN                                         & 46.11          & 42.05          & 34.76          \\ \hline
DnCNN + SAN                                      & 45.22          & 39.58          & 32.73          \\ \hline
DnCNN + RCAN                                     & 41.25          & 35.30           & 27.40           \\ \hline
DnCNN + USRNet                                   & 43.16          & 37.12          & 30.61          \\ \hline
USRNet                                           & 43.32          & 37.43          & 30.63          \\ \hline
USRNet + NRIBP                                   & 41.92          & 36.34          & 29.54          \\ \hline
Our tiny model                                   & \textit{46.91} & \textit{42.93} & \textit{36.71} \\ \hline
Our full model                                   & \textbf{46.98} & \textbf{43.33} & \textbf{37.54} \\ \hline
\end{tabular}}
\label{quan_real}
\end{table}
\begin{figure*}[ht]
\begin{minipage}{1\linewidth}
\centerline{\includegraphics[scale=0.6]{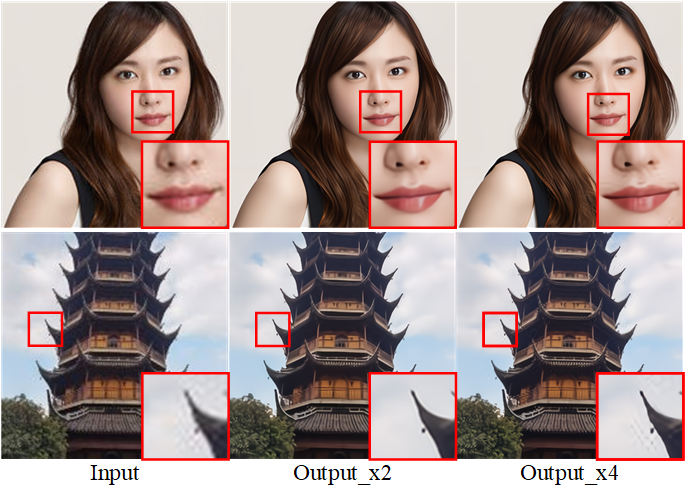}}
\end{minipage}
\centering
\caption{Two failure cases on the collected avatar images dataset. Rescaling all of these images for better comparison.}
\label{fig:fail}
\end{figure*} 

\subsection{Super-resolving Images from Social Media}
\label{real_app}
Social media platforms such as \textit{WeChat} have become popular for internet users to share photos. However, images uploaded by users would be downscaled and compressed by the service due to the limitations of storage capacity and transmission bandwidth. For example, users’ avatar images in \textit{WeChat} are downscaled and compressed, and their degradation histories are unknown to users. That is, the down-sampling kernels and compression parameters are not available. As a real-world application, we apply our trained models directly to super-resolve these severely scaled and compressed avatar images to demonstrate the generalization ability of our model. We have constructed a dataset containing 50 \textit{WeChat} avatar images. Specifically, 5 randomly selected \textit{WeChat} users (3 females and 2 males), and each user has volunteered to provide 10 avatar images from their friend lists after obtaining approval from the content owners. We evaluate the performance of the proposed method on this dataset and compare it with the competitors in Section \ref{compare}. In this real-world testing, all the SISR models, including our tiny and full models, remain the same as the ones described in Sections \ref{dataset} and \ref{compare}, without any retraining or fine-tuning to fit the dataset. We simply up-scale these 50 images by a factor of 2, 3 and 4 respectively. Due to the lack of ground truths, a state-of-the-art blind quality assessment method, abbreviated as CaHDC \cite{ref79}, is utilized for quantitative evaluations. Higher score means better image quality for this assessment method. The average values of CaHDC are shown in Table \ref{quan_real}, in which the best performance is highlighted in bold, and the second-best results are distinguished by italics. Three visual examples are further provided in Fig. \ref{fig:real}. From Table \ref{quan_real} and Fig. \ref{fig:real}, we can see that our method has much better generalization ability for real-world applications. More details about the collected avatar dataset, as well as more visual examples, can be found in our online materials \cite{ref73}.

However, super-resolving real-world images whose formation remains unknown is very challenging. Thus, there are also some failure cases when applying our model to these real-world images. As shown in Fig. \ref{fig:fail}, it can be seen that, some artefacts might be enlarged if we super-resolve the input with the scale factor 4 by our model, although our results for up-scaling factor 2 are satisfactory. It may be attributed to the fact that the model trained on the data with scale factor of 4 would struggle to enhance weak edges. One possible solution to this limitation is to make use of the up-scaling factor as another input to provide auxiliary information for the adaptive combination in our framework.

\section{Concluding Remarks}
In this paper, we propose a parallel and series integration framework to super-resolve compressed LR images. This framework includes two modules: the ARM and the REM. Both modules are based on deep neural networks and share similar network architectures. Between the ARM and the REM, both parallel and series flows are included. On one hand, compressed LR image is received and processed by the ARM and the REM in a parallel way. As a result, the original information in the input is fully available for the both modules without any loss or change. On the other hand, two series flows are formed by regarding the output of one module as the auxiliary input of the other one. These series streams enable the information exchange between the ARM and the REM so that the two modules can facilitate each other. In this way, our framework is capable of super-resolving compressed images without assuming apriori compressions.
Furthermore, to make better use of the auxiliary inputs, a modified non-local operator and an adaptive combination module, both with learnable parameters, are introduced, which have helped improving performances. Experiments are conducted on the LR images with various compression configurations. Extensive comparisons on both benchmark dataset and social media dataset demonstrate the advantage of the proposed model over state-of-the-art SISR models.

\section{Acknowledgements}
\label{sec:Acknowledgements}

This work was in part by Guangdong Basic and Applied Basic Research Foundation with No.2021A1515011584, and the Shenzhen Research and Development Program under Grant JCYJ20220531102408020 and Grant JCYJ-20200109105008228, and in part by National Natural Science Foundation of China under Grant 62271323.


\bibliography{mybibfile}

\begin{thebibliography}{10}
\expandafter\ifx\csname url\endcsname\relax
  \def\url#1{\texttt{#1}}\fi
\expandafter\ifx\csname urlprefix\endcsname\relax\def\urlprefix{URL }\fi
\expandafter\ifx\csname href\endcsname\relax
  \def\href#1#2{#2} \def\path#1{#1}\fi

\bibitem{ref1}
C.-Y. Yang, C.~Ma, M.-H. Yang, Single-image super-resolution: A benchmark, in:
  Proceedings of European Conference on Computer Vision, 2014.

\bibitem{ref3}
W.~W.~W. Zou, P.~C. Yuen, Very low resolution face recognition problem, IEEE
  Transactions on Image Processing 21~(1) (2012) 327--340.

\bibitem{ref4}
Z.~Xiong, X.~Sun, F.~Wu, Super-resolution for low quality thumbnail images, in:
  2008 IEEE International Conference on Multimedia and Expo, 2008, pp.
  181--184.

\bibitem{ref31}
Z.~Xiong, X.~Sun, F.~Wu, Robust web image/video super-resolution, IEEE
  Transactions on Image Processing 19~(8) (2010) 2017--2028.

\bibitem{ref32}
L.-W. Kang, C.-C. Hsu, B.~Zhuang, C.-W. Lin, C.-H. Yeh, Learning-based joint
  super-resolution and deblocking for a highly compressed image, IEEE
  Transactions on Multimedia 17~(7) (2015) 921--934.

\bibitem{ref33}
O.-Y. Lee, J.-W. Lee, D.-Y. Lee, J.-O. Kim, Joint super-resolution and
  compression artifact reduction based on dual-learning, in: 2016 Visual
  Communications and Image Processing (VCIP), 2016, pp. 1--4.

\bibitem{ref34}
T.~Li, X.~He, L.~Qing, Q.~Teng, H.~Chen, An iterative framework of cascaded
  deblocking and superresolution for compressed images, IEEE Transactions on
  Multimedia 20~(6) (2018) 1305--1320.

\bibitem{ref35}
H.~Chen, X.~He, C.~Ren, L.~Qing, Q.~Teng, Cisrdcnn: Super-resolution of
  compressed images using deep convolutional neural networks, Neurocomputing
  285 (2018) 204--219.

\bibitem{ref36}
A.~R. Zamir, T.-L. Wu, L.~Sun, W.~B. Shen, B.~E. Shi, J.~Malik, S.~Savarese,
  Feedback networks, in: Proceedings of the IEEE conference on computer vision
  and pattern recognition, 2017, pp. 1308--1317.

\bibitem{ref15}
J.~Kim, J.~K. Lee, K.~M. Lee, Accurate image super-resolution using very deep
  convolutional networks, in: 2016 IEEE Conference on Computer Vision and
  Pattern Recognition (CVPR), 2016, pp. 1646--1654.

\bibitem{ref37}
K.~He, X.~Zhang, S.~Ren, J.~Sun, Deep residual learning for image recognition,
  in: Proceedings of the IEEE conference on computer vision and pattern
  recognition, 2016, pp. 770--778.

\bibitem{ref5}
W.~Freeman, T.~Jones, E.~Pasztor, Example-based super-resolution, IEEE Computer
  Graphics and Applications 22~(2) (2002) 56--65.

\bibitem{ref7}
H.~Chang, D.-Y. Yeung, Y.~Xiong, Super-resolution through neighbor embedding,
  in: Proceedings of the 2004 IEEE Computer Society Conference on Computer
  Vision and Pattern Recognition, 2004. CVPR 2004., Vol.~1, 2004, pp. I--I.

\bibitem{ref8}
J.~Yang, J.~Wright, T.~S. Huang, Y.~Ma, Image super-resolution via sparse
  representation, IEEE Transactions on Image Processing 19~(11) (2010)
  2861--2873.

\bibitem{ref9}
R.~Timofte, V.~De~Smet, L.~Van~Gool, A+: Adjusted anchored neighborhood
  regression for fast super-resolution, in: Asian conference on computer
  vision, Springer, 2014, pp. 111--126.

\bibitem{ref13}
C.~Dong, C.~C. Loy, K.~He, X.~Tang, Image super-resolution using deep
  convolutional networks, IEEE transactions on pattern analysis and machine
  intelligence 38~(2) (2015) 295--307.

\bibitem{ref16}
B.~Lim, S.~Son, H.~Kim, S.~Nah, K.~Mu~Lee, Enhanced deep residual networks for
  single image super-resolution, in: Proceedings of the IEEE conference on
  computer vision and pattern recognition workshops, 2017, pp. 136--144.

\bibitem{ref19}
Y.~Zhang, K.~Li, K.~Li, L.~Wang, B.~Zhong, Y.~Fu, Image super-resolution using
  very deep residual channel attention networks, in: Proceedings of the
  European conference on computer vision (ECCV), 2018, pp. 286--301.

\bibitem{ref20}
T.~Dai, J.~Cai, Y.~Zhang, S.-T. Xia, L.~Zhang, Second-order attention network
  for single image super-resolution, in: Proceedings of the IEEE/CVF Conference
  on Computer Vision and Pattern Recognition, 2019, pp. 11065--11074.

\bibitem{ref24}
Z.~Li, J.~Yang, Z.~Liu, X.~Yang, G.~Jeon, W.~Wu, Feedback network for image
  super-resolution, in: Proceedings of the IEEE/CVF Conference on Computer
  Vision and Pattern Recognition, 2019, pp. 3867--3876.

\bibitem{ref25}
C.~Ren, X.~He, T.~Q. Nguyen, Single image super-resolution via adaptive
  high-dimensional non-local total variation and adaptive geometric feature,
  IEEE Transactions on Image Processing 26~(1) (2017) 90--106.

\bibitem{ref26}
Z.-S. Liu, L.-W. Wang, C.-T. Li, W.-C. Siu, Y.-L. Chan, Image super-resolution
  via attention based back projection networks, in: 2019 IEEE/CVF International
  Conference on Computer Vision Workshop (ICCVW), IEEE, 2019, pp. 3517--3525.

\bibitem{ref11}
W.~Yang, X.~Zhang, Y.~Tian, W.~Wang, J.-H. Xue, Q.~Liao, Deep learning for
  single image super-resolution: A brief review, IEEE Transactions on
  Multimedia 21~(12) (2019) 3106--3121.

\bibitem{ref28}
T.~Köhler, M.~Bätz, F.~Naderi, A.~Kaup, A.~Maier, C.~Riess, Toward bridging
  the simulated-to-real gap: Benchmarking super-resolution on real data, IEEE
  Transactions on Pattern Analysis and Machine Intelligence 42~(11) (2020)
  2944--2959.

\bibitem{ref29}
A.~Singh, F.~Porikli, N.~Ahuja, Super-resolving noisy images, in: Proceedings
  of the IEEE Conference on Computer Vision and Pattern Recognition, 2014, pp.
  2846--2853.

\bibitem{ref39}
S.~Huang, J.~Sun, Y.~Yang, Y.~Fang, P.~Lin, Y.~Que, Robust single-image
  super-resolution based on adaptive edge-preserving smoothing regularization,
  IEEE Transactions on Image Processing 27~(6) (2018) 2650--2663.

\bibitem{ref40}
J.-S. Yoo, J.-O. Kim, Noise-robust iterative back-projection, IEEE Transactions
  on Image Processing 29 (2020) 1219--1232.

\bibitem{ref41}
K.~Zhang, W.~Zuo, L.~Zhang, Learning a single convolutional super-resolution
  network for multiple degradations, in: 2018 IEEE/CVF Conference on Computer
  Vision and Pattern Recognition, 2018, pp. 3262--3271.

\bibitem{ref42}
K.~Zhang, W.~Zuo, L.~Zhang, Deep plug-and-play super-resolution for arbitrary
  blur kernels, in: 2019 IEEE/CVF Conference on Computer Vision and Pattern
  Recognition (CVPR), 2019, pp. 1671--1681.

\bibitem{ref43}
K.~Zhang, L.~Van~Gool, R.~Timofte, Deep unfolding network for image
  super-resolution, in: 2020 IEEE/CVF Conference on Computer Vision and Pattern
  Recognition (CVPR), 2020, pp. 3214--3223.

\bibitem{ref30}
J.~Cai, H.~Zeng, H.~Yong, Z.~Cao, L.~Zhang, Toward real-world single image
  super-resolution: A new benchmark and a new model, in: 2019 IEEE/CVF
  International Conference on Computer Vision (ICCV), 2019, pp. 3086--3095.

\bibitem{ref45}
Y.~Zhang, S.~Liu, C.~Dong, X.~Zhang, Y.~Yuan, Multiple cycle-in-cycle
  generative adversarial networks for unsupervised image super-resolution, IEEE
  Transactions on Image Processing 29 (2020) 1101--1112.

\bibitem{ref44}
X.~Hu, Z.~Zhang, C.~Shan, Z.~Wang, L.~Wang, T.~Tan, Meta-usr: A unified
  super-resolution network for multiple degradation parameters, IEEE
  Transactions on Neural Networks and Learning Systems (2020) 1--15.

\bibitem{ref46}
J.~Apostolopoulos, N.~Jayant, Postprocessing for very low bit-rate video
  compression, IEEE Transactions on Image Processing 8~(8) (1999) 1125--1129.

\bibitem{ref47}
D.~Sun, W.-K. Cham, Postprocessing of low bit-rate block dct coded images based
  on a fields of experts prior, IEEE Transactions on Image Processing 16~(11)
  (2007) 2743--2751.

\bibitem{ref48}
G.~Triantafyllidis, D.~Tzovaras, M.~Strintzis, Blocking artifact detection and
  reduction in compressed data, IEEE Transactions on Circuits and Systems for
  Video Technology 12~(10) (2002) 877--890.

\bibitem{ref49}
A.~Foi, V.~Katkovnik, K.~Egiazarian, Pointwise shape-adaptive dct for
  high-quality denoising and deblocking of grayscale and color images, IEEE
  Transactions on Image Processing 16~(5) (2007) 1395--1411.

\bibitem{ref50}
C.~Dong, Y.~Deng, C.~C. Loy, X.~Tang, Compression artifacts reduction by a deep
  convolutional network, in: 2015 IEEE International Conference on Computer
  Vision (ICCV), 2015, pp. 576--584.

\bibitem{ref53}
K.~Zhang, W.~Zuo, Y.~Chen, D.~Meng, L.~Zhang, Beyond a gaussian denoiser:
  Residual learning of deep cnn for image denoising, IEEE Transactions on Image
  Processing 26~(7) (2017) 3142--3155.

\bibitem{ref55}
X.~Fu, Z.-J. Zha, F.~Wu, X.~Ding, J.~Paisley, Jpeg artifacts reduction via deep
  convolutional sparse coding, in: 2019 IEEE/CVF International Conference on
  Computer Vision (ICCV), 2019, pp. 2501--2510.
\newblock \href {http://dx.doi.org/10.1109/ICCV.2019.00259}
  {\path{doi:10.1109/ICCV.2019.00259}}.

\bibitem{ref56}
J.~Liu, D.~Liu, W.~Yang, S.~Xia, X.~Zhang, Y.~Dai, A comprehensive benchmark
  for single image compression artifact reduction, IEEE Transactions on Image
  Processing 29 (2020) 7845--7860.

\bibitem{ref57}
Y.~Chen, T.~Pock, Trainable nonlinear reaction diffusion: A flexible framework
  for fast and effective image restoration, IEEE Transactions on Pattern
  Analysis and Machine Intelligence 39~(6) (2017) 1256--1272.

\bibitem{ref58}
H.~Chang, M.~K. Ng, T.~Zeng, Reducing artifacts in jpeg decompression via a
  learned dictionary, IEEE Transactions on Signal Processing 62~(3) (2014)
  718--728.

\bibitem{ref59}
C.~Wang, J.~Zhou, S.~Liu, Adaptive non-local means filter for image deblocking,
  Signal Processing: Image Communication 28~(5) (2013) 522--530.

\bibitem{ref60}
Y.~Zhang, K.~Li, K.~Li, B.~Zhong, Y.~Fu, Residual non-local attention networks
  for image restoration, in: International Conference on Learning
  Representations, 2019.

\bibitem{ref18}
Y.~Zhang, Y.~Tian, Y.~Kong, B.~Zhong, Y.~Fu, Residual dense network for image
  super-resolution, in: Proceedings of the IEEE conference on computer vision
  and pattern recognition, 2018, pp. 2472--2481.

\bibitem{ref80}
Y.~Hu, X.~Gao, J.~Li, Y.~Huang, H.~Wang, Single image super-resolution with
  multi-scale information cross-fusion network, Signal Processing 179 (2021)
  107831.

\bibitem{ref61}
G.~Wallace, The jpeg still picture compression standard, IEEE Transactions on
  Consumer Electronics 38~(1) (1992) xviii--xxxiv.

\bibitem{ref68}
Y.~Bengio, J.~Louradour, R.~Collobert, J.~Weston, Curriculum learning, in:
  Proceedings of the 26th annual international conference on machine learning,
  2009, pp. 41--48.

\bibitem{ref63}
M.~S.~M. Sajjadi, R.~Vemulapalli, M.~Brown, Frame-recurrent video
  super-resolution, in: 2018 IEEE/CVF Conference on Computer Vision and Pattern
  Recognition, 2018, pp. 6626--6634.

\bibitem{ref64}
A.~Buades, B.~Coll, J.-M. Morel, Nonlocal image and movie denoising,
  International journal of computer vision 76~(2) (2008) 123--139.

\bibitem{ref65}
X.~Wang, R.~Girshick, A.~Gupta, K.~He, Non-local neural networks, in:
  Proceedings of the IEEE conference on computer vision and pattern
  recognition, 2018, pp. 7794--7803.

\bibitem{ref66}
E.~Lesellier, J.~Jung, Robust wavelet-based arbitrary grid detection for mpeg,
  in: Proceedings. International Conference on Image Processing, Vol.~3, 2002,
  pp. III--III.

\bibitem{ref67}
K.~He, X.~Zhang, S.~Ren, J.~Sun, Identity mappings in deep residual networks,
  in: European conference on computer vision, Springer, 2016, pp. 630--645.

\bibitem{ref70}
G.~Developers, \href{http://r0k.us/graphics/kodak/}{Webp - a new image format
  for the web.} (2013).
\newline\urlprefix\url{http://r0k.us/graphics/kodak/}

\bibitem{ref71}
R.~Timofte, E.~Agustsson, L.~Van~Gool, M.-H. Yang, L.~Zhang, Ntire 2017
  challenge on single image super-resolution: Methods and results, in:
  Proceedings of the IEEE conference on computer vision and pattern recognition
  workshops, 2017, pp. 114--125.

\bibitem{ref69}
D.~P. Kingma, J.~Ba, Adam: {A} method for stochastic optimization, in:
  Y.~Bengio, Y.~LeCun (Eds.), 3rd International Conference on Learning
  Representations, {ICLR} 2015, San Diego, CA, USA, May 7-9, 2015, Conference
  Track Proceedings, 2015.

\bibitem{ref72}
R.~Franzen, \href{https://developers.google.com/speed/webp/}{Kodak lossless
  true color image suite} (2010).
\newline\urlprefix\url{https://developers.google.com/speed/webp/}

\bibitem{ref73}
H.~Luo, \href{http://www.vista.ac.cn/cisr-pcs/}{Online materials for
  super-resolving compressed images via parallel and series integration of
  artifact reduction and resolution enhancement.} (2021).
\newline\urlprefix\url{http://www.vista.ac.cn/cisr-pcs/}

\bibitem{ref74}
X.~Liu, M.~Tanaka, M.~Okutomi, Single-image noise level estimation for blind
  denoising, IEEE Transactions on Image Processing 22~(12) (2013) 5226--5237.

\bibitem{ref75}
S.~Liu, Q.~Liao, J.-H. Xue, F.~Zhou, Defocus map estimation from a single image
  using improved likelihood feature and edge-based basis, Pattern Recognition
  107 (2020) 107485.

\bibitem{ref76}
Z.~Wang, A.~Bovik, H.~Sheikh, E.~Simoncelli, Image quality assessment: from
  error visibility to structural similarity, IEEE Transactions on Image
  Processing 13~(4) (2004) 600--612.

\bibitem{ref77}
H.~Sheikh, A.~Bovik, G.~de~Veciana, An information fidelity criterion for image
  quality assessment using natural scene statistics, IEEE Transactions on Image
  Processing 14~(12) (2005) 2117--2128.

\bibitem{ref78}
F.~Zhou, R.~Yao, B.~Liu, G.~Qiu, Visual quality assessment for super-resolved
  images: Database and method, IEEE Transactions on Image Processing 28~(7)
  (2019) 3528--3541.

\bibitem{ref79}
J.~Wu, J.~Ma, F.~Liang, W.~Dong, G.~Shi, W.~Lin, End-to-end blind image quality
  prediction with cascaded deep neural network, IEEE Transactions on Image
  Processing 29 (2020) 7414--7426.

\end{thebibliography}

\end{document}